\documentclass[%
 reprint,
superscriptaddress,
nofootinbib,
 amsmath,amssymb,
 onecolumn,
prd,
]{revtex4-2}

\usepackage{graphicx}% Include figure files
\usepackage{dcolumn}% Align table columns on decimal point
\usepackage{bm}% bold math
\usepackage[colorlinks]{hyperref}% add hypertext capabilities
\usepackage{xcolor}
\usepackage[caption=false]{subfig}
\newcommand{\bv}[1]{\boldsymbol{#1}}

\newcommand{\Mpch}{h^{-1}\mathrm{Mpc}}

\newcommand{\delD}[1]{(2\pi)^3\delta_\mathrm{D}\left({#1}\right)}

\newcommand{\av}[1]{\left\langle{#1}\right\rangle} 

\newcommand{\vk}{\bv k}

\newcommand{\vx}{\bv x}

\renewcommand{\vr}{\bv r}

\newcommand{\G}{\mathcal{G}}

\renewcommand{\P}{\mathcal{P}}

\newcommand{\hn}{\hat{\bv n}}
\newcommand{\hz}{\hat{\bv z}}
\newcommand{\hk}{\hat{\bv k}}

\newcommand{\tj}[6]{\begin{pmatrix} {#1} & {#2} & {#3}\\ {#4} & {#5} & {#6}\end{pmatrix}}

\usepackage{color}
\def\beq{\begin{eqnarray}}
\def\eeq{\end{eqnarray}}

\DeclareSymbolFont{toneletters}{T1}{\familydefault}{m}{it}
\SetSymbolFont{toneletters}{bold}{T1}{\familydefault}{bx}{it}

\DeclareMathSymbol\edth{\mathord}{toneletters}{"F0}

\usepackage{empheq}

\definecolor{darkgreen}{RGB}{0,120,0}
\definecolor{brown}{RGB}{120,60,0}

\newcommand{\resub}[1]{#1}%\textcolor{darkgreen}{#1}}

\begin{document}

%\preprint{APS/123-QED}

%\title{What Can We Learn from Combining Galaxy Lensing and the Polarized Sunyaev-Zel'dovich Effect?}% Force line breaks with \\
%\thanks{A footnote to the article title}%
\title{Novel Cosmological Tests from Combining Galaxy Lensing and the Polarized Sunyaev-Zel'dovich Effect}
\author{Oliver H.\,E. Philcox}
\email{ohep2@cantab.ac.uk}
\affiliation{Department of Astrophysical Sciences, Princeton University, Princeton, NJ 08540, USA}%
\affiliation{School of Natural Sciences, Institute for Advanced Study, 1 Einstein Drive, Princeton, NJ 08540, USA}
\affiliation{Center for Theoretical Physics, Department of Physics,
Columbia University, New York, NY 10027, USA}
\affiliation{Simons Society of Fellows, Simons Foundation, New York, NY 10010, USA}
\author{Matthew C. Johnson}
\email{mjohnson@perimeterinstitute.ca}
\affiliation{Perimeter Institute for Theoretical Physics, 31 Caroline St N, Waterloo, ON N2L 2Y5, Canada}
\affiliation{Department of Physics and Astronomy, York University, Toronto, ON M3J 1P3, Canada}

%\collaboration{MUSO Collaboration}%\noaffiliation

%\date{\today}% It is always \today, today,
             %  but any date may be explicitly specified

\begin{abstract}
The polarized Sunyaev-Zel'dovich (pSZ) effect is sourced by the Thomson scattering of CMB photons from distant free electrons and yields a novel view of the CMB quadrupole throughout the observable Universe. Galaxy shear measures the shape distortions of galaxies, probing both their local environment and the intervening matter distribution. Both observables have been shown to give interesting constraints on the cosmological model; in this work we ask: what can be learnt from their combination? The pSZ-shear cross-spectrum measures the shear-galaxy-polarization bispectrum (\textit{i.e.}\ $\av{\gamma \delta_g(Q\pm iU)}$) and contains contributions from three main phenomena: (1) the Sachs-Wolfe (SW) effect, (2) the integrated Sachs-Wolfe (ISW) effect, (3) inflationary gravitational waves. Since the modes contributing to the pSZ signal are not restricted to the Earth's past lightcone, the low-redshift cross-spectra could provide a novel constraint on dark energy properties via the ISW effect, whilst the SW signal is sourced by a coupling of scalar modes at very different times (recombination and the lensing redshift), but at similar positions; this provides a unique probe of the Universe's homogeneous time evolution. We give expressions for all major contributions to the \resub{galaxy} shear, galaxy \resub{density}, and pSZ auto- and cross-spectra, and evaluate their detectability via Fisher forecasts. Despite significant theoretical utility, the \resub{shear} cross-spectra will be challenging to detect: combining CMB-S4 with the Rubin observatory yields a $1.6\sigma$ detection of the ISW contribution, though this increases to $5.2\sigma$ for a futuristic experiment involving CMB-HD and a higher galaxy sample density. For parity-even (parity-odd) tensors, we predict a $1\sigma$ limit of $\sigma(r) = 0.9$ ($0.2$) for CMB-S4 and Rubin, or $0.3$ ($0.06$) for the more futuristic setup. Whilst this is significantly better than the constraints from galaxy shear alone (and contains fewer systematics than most auto-spectra), it is unlikely to be competitive, but may serve as a useful cross-check. 
\end{abstract}

%\keywords{Suggested keywords}%Use showkeys class option if keyword
                              %display desired
\maketitle

%\tableofcontents

%\section{Introduction}\label{sec: intro}

%\vskip 4pt

\section{Introduction \& Motivation}

Cosmology exists on the lightcone. Almost all cosmological observables follow the paths of photon geodesics from their source to us, whether they be from distant galaxies or the cosmic microwave background; as such, our knowledge of the Universe is restricted to sections of the light cone. With conventional measurements,  our knowledge is limited to the surface of the light-cone, rather than its interior. One consequence of this is that  standard cosmological observables cannot directly test cosmological homogeneity; rather, they can probe only anisotropies and time evolution  \citep[e.g.,][]{Maartens:2011yx}.

\begin{figure}
    \centering
    \subfloat[Tensors and the Integrated Sachs-Wolfe Effect]{%
      \includegraphics[width=0.48\textwidth]{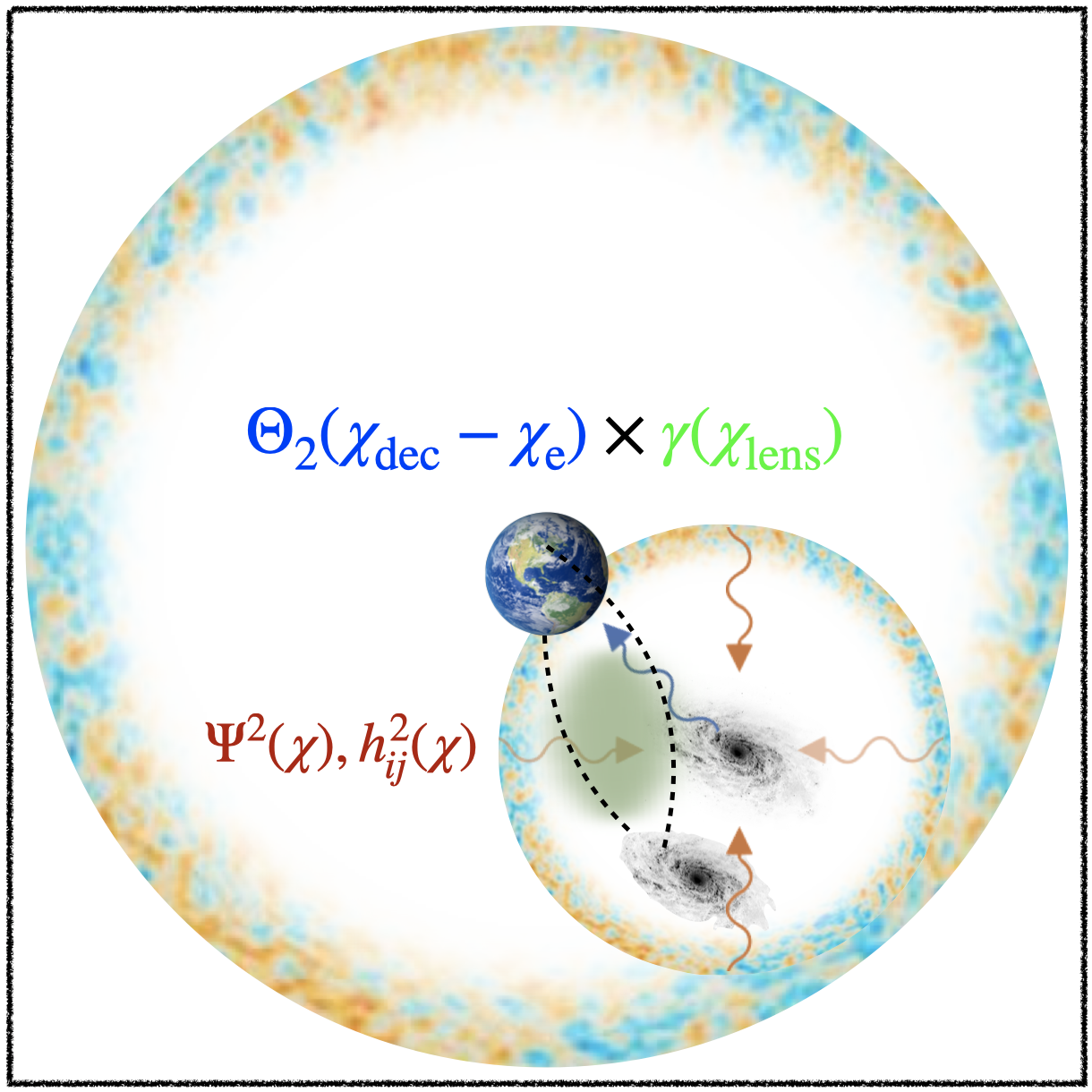}\label{subfig: cartoon1}
    }
    \hfill
    \subfloat[Unequal Time Sachs-Wolfe Effect]{%
      \includegraphics[width=0.48\textwidth]{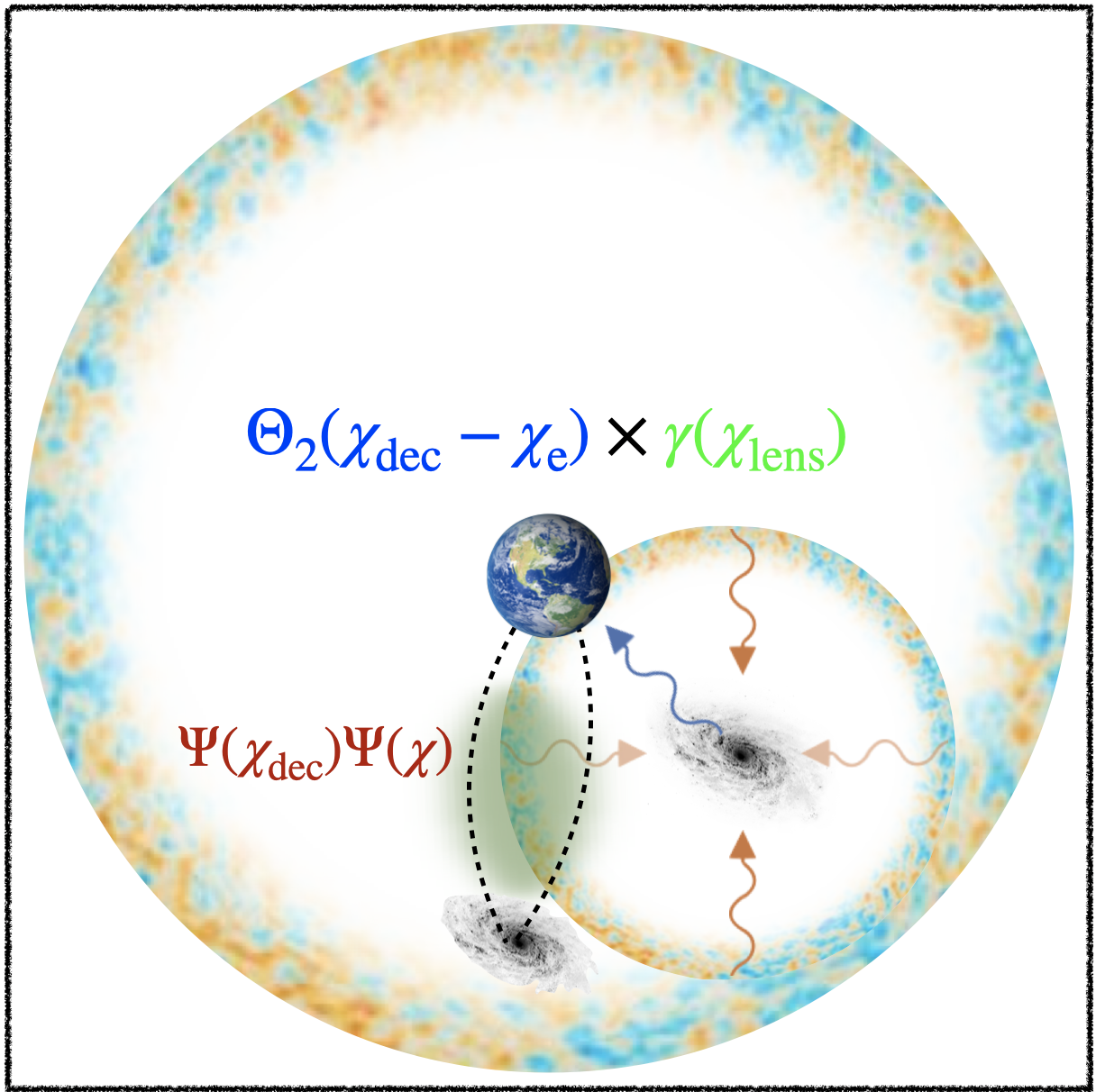}\label{subfig: cartoon2}
    }
    \caption{Cartoon of the pSZ-shear cross-correlations considered in this work. The pSZ effect is sourced by the CMB observed at some distant galaxy (brown arrows), whose quadrupole moment, $\Theta_2$, is anisotropically scattered (blue arrow) and reaches the Earth. This is sensitive to physics on the lightcone of the distant galaxy, and that between the distant galaxy and the Earth. In contrast, shear is sourced by the perturbations to a photon geodesic imprinted by scalars, $\Psi$, and tensors, $h_{ij}$, as it traverses its worldline on the Earth's lightcone (shown in black dotted lines). If the lens (green region) lies close to the scattering galaxy, a correlation will be induced due to both sets of photons (scattered CMB and weak lensing) experiencing the same scalar (through the integrated Sachs-Wolfe effect) and tensor modes. If the lens lies further than the pSZ-source, other correlations can arise, such as one between the scalar potential at last scattering, $\Psi(\chi_{\rm dec})$ and that sourcing gravitational lensing. Unlike most physics observables, this correlates two effects at very different times but at the same physical location.}\label{fig: cartoon}
\end{figure}

The kinetic and polarized Sunyaev-Zel'dovich (SZ) effects are different in this respect. Both are caused by the scattering of cosmic microwave background (CMB) photons from a galaxy at some radial comoving distance $\chi_e$ from the observer. Importantly, this CMB is not the same as the one observed by us today. Instead, the galaxy scatters the locally-observed CMB, which is sourced by the {\em interior} of our past light cone as depicted in Fig.\,\ref{fig: cartoon}. If the local CMB is anisotropic, we will observe a signature from the direction of the galaxy. For the kinetic SZ (kSZ) effect \citep[e.g.,][]{Zhang:2015uta,Terrana:2016xvc,Sunyaev:1980nv,Audit:1998uk,Challinor:1999yz,Shao:2010md,Smith:2018bpn}, this is caused by the CMB dipole observed by the galaxy, whose dominant component of the dipole is the galaxy's peculiar velocity. For the polarized SZ (pSZ) effect \citep[e.g.,][]{Sazonov:1999zp,Sunyaev:1980nv,Deutsch:2017ybc,Deutsch:2018umo,Deutsch:2017cja,Itoh:1998wv,Emritte:2016xns,Alizadeh:2012vy,Kamionkowski:1997na,Hall:2014wna,Seto:2000uc,Abramo:2006gp,Bunn:2006mp,Liu:2016fqc,Portsmouth:2004mk,Baumann:2003xb,Seto:2005de,Pan:2019dax,Hotinli:2022wbk}, this is instead a consequence of Thomson scattering caused by the local CMB quadrupole. Such effects propagate to a distortion in the CMB temperature and polarization anisotropies measured on Earth that correlate with the galaxy density at the scattering location; by utilizing this correlation, one can extract the dipole and quadrupole {\em fields} as a function of position and distance \citep[e.g.,][]{Deutsch:2017ybc,Kamionkowski:1997na,Shao:2010md,Smith:2018bpn,Abramo:2006gp,Deutsch:2017cja}. Crucially, this is a measurement of two things: (a) the CMB primaries and secondaries observed at the source at $\chi_e$ (orange arrow in Fig.\,\ref{fig: cartoon}) and (b) the distribution of matter between $\chi_e$ to the observer at $\chi=0$ (blue arrow in Fig.\,\ref{fig: cartoon}). Only the latter quantity is constrained to lie \textit{on} the Earth's lightcone; the former lies instead \textit{within} it, as a result of photons taking a non-direct route to the Earth, and is the subject of interest in this work.

Cosmic shear is a key observable in twenty-first century cosmology. This measures the shape distortions of galaxies as a function of position and distance, which carries information both about the galaxies' local environments (via the `intrinsic alignment' mechanism, \citep[e.g.,][]{2004PhRvD..70f3526H}) and the intervening spacetime (via gravitational lensing, \citep[e.g.,][]{2001PhR...340..291B}). In contrast to kSZ and pSZ these quantities lie on the lightcone (or at least on its first-order perturbations); however, they are sensitive to a number of interesting features. Conventionally, cosmic shear is used to probe the integrated matter density from a source at some distance $\chi_{\rm lens}$ from the observer (located at $\chi=0$), and has been shown to give tight constraints on the matter density and clustering amplitude \citep[e.g.,][]{2021arXiv210513549D}. Noting that the matter density is nothing but a gauge transform of the scalar metric potential, one may ask whether galaxy shapes are sensitive to other metric perturbations, such as gravitational waves. As shown in a number of works \citep[e.g.,][]{2008PhRvD..77j3515S,2020JCAP...07..005B,2012PhRvD..86h3513S,2014PhRvD..89h3507S} such an effect does exist, and contributes both to lensing and intrinsic alignment. This occurs since the galaxy shape is a tensorial (spin-two) observable, and thus can couple to tensor metric perturbations. Unfortunately, the size of such an effect is generally small, since gravitational waves significantly decay after inflation, and, moreover, the principal observable, the shear $B$-mode, is usually discarded on the grounds of systematics or just used for null tests \citep[e.g.,][]{2021arXiv210513549D}. As such, most efforts to measure gravitational waves have been directed towards the primary CMB.

The next decade will yield unprecedented volumes of cosmological data, both from the CMB, due to experiments such as the Simons Observatory \citep{SimonsObservatory:2018koc} and CMB-S4 \citep{Abazajian:2019tiv}, and large-scale structure (LSS), with photometric surveys such as Rubin (hereafter VRO) \citep{LSSTScience:2009jmu} and Euclid, as well as spectroscopic instruments including DESI \citep{DESI:2016fyo} and MegaMapper \citep{Schlegel:2019eqc}. The incoming avalanche motivates us to consider new ways of probing the Universe, in particular those constraining hitherto poorly understood degrees of freedom. In this work, we will add to such an effort by considering the detectability and utility of cross-correlations between the pSZ contribution to the CMB polarization anisotropies and cosmic shear   (\textit{i.e.}\ a $\av{\gamma g(Q\pm iU)}$ three-point function). The physical consequences of each observable has been considered in the past \citep[e.g.,][]{Deutsch:2018umo,Pan:2019dax,Seto:2000uc,Seto:2005de,2012PhRvD..86h3513S,2001PhR...340..291B,2020PASJ...72...16H,2021arXiv210513549D}, however the correlations described in Fig.~\ref{fig: cartoon} have  yet to be assessed. Performing analyses using cross-spectra can be particularly enlightening since (a) they are often less sensitive to systematic effects than auto-spectra, (b) incomplete correlations can allow specific, and interesting, pieces of the signal to be extracted. That said, a strong correlation is needed for an observable to be useful, and it is unclear, \textit{a priori}, whether pSZ and shear (or indeed, pSZ and galaxy positions) satisfy this.

There are two sources of scalar pSZ-shear correlations. The first occurs when the observables probe matter in the same region of space at similar times, as in Fig.\,\ref{subfig: cartoon1}. An important contributor to the pSZ signal is the Integrated Sachs-Wolfe (ISW) effect \citep[e.g.,][]{Baumann:2003xb,Seto:2005de}, which probes the Universe in the vicinity of the scattering galaxy at $\chi_e$; for shear, photons emanating from the galaxy at $\chi_{\rm lens}\gtrsim \chi_e$ are lensed by the same matter distribution.\footnote{Strictly, such a correlation can be sourced also by the lensing of pSZ photons after their scattering; this phenomena is second-order however, and likely to be small.} In this way, the pSZ-shear correlation probes the local potential (or rather its time derivative), via the ISW effect, and, unlike detections obtained from the primary CMB is not limited to the Earth's lightcone. 

The second possibility is to have correlations between spatially-close regions of the Universe at vastly different times. This principally occurs for $\chi_{\rm lens}>\chi_e$, whence the gravitational potential sourcing lensing (at a time $\chi<\chi_e$) is also the source of Sachs-Wolfe (SW) effects at the local last-scattering-surface seen (after rescattering) in the pSZ effect. Since pSZ is not restricted to the lightcone, this scenario is fully permissible (cf.\,Fig.\,\ref{subfig: cartoon2}) and arises since the scattering photon does not take a direct path from the redshift of decoupling until today. Mathematically, the phenomena is caused by a correlation of the SW potential $\Psi(\vx,\chi_{\rm dec})$ and the lensing potential $\Psi(\vx',\chi)$ at large relative time ($\chi\ll \chi_{\rm dec}$) but small relative position ($|\vx-\vx'| \ll \chi$). The ability to correlate potentials at such different times is particularly unusual in cosmology (and made possible only via the off-lightcone effects), and its detection would certainly be of great interest. If measured, this would allow one to probe the local growth function $D(\chi,\vx)$ at two redshifts simultaneously, and, in principle, allow for a spatially resolved map of $D(\chi,\vx)/D(\chi_{\rm dec},\vx)$, given additional geometric information (cf.\,\S\ref{sec:cosmotests}).

Gravitational waves can also be probed using the cross-correlation of pSZ and shear. The intuition for this is straightforward: both pSZ and shear measure tensorial quantities, the CMB quadrupole and the galaxy shape tensor $\gamma_{ij}$. For pSZ, tensor signatures (of both odd and even parity) arise in the same manner as the primary CMB: predominantly from the gravitational effects imparted on radiation in the time after recombination, whilst for shear, this is sourced by lensing and intrinsic effects. While  gravitational wave signatures in shear are very weak \citep{2012PhRvD..86h3513S,2014PhRvD..89h3507S,2008PhRvD..77j3515S,2020JCAP...07..005B}, they have been shown to be observable in the pSZ signal accessible to future experiments \citep{Alizadeh:2012vy,Deutsch:2018umo,Deutsch:2017ybc}, thus it is interesting to consider whether their cross-correlation can be of use, and whether the pSZ can be used to boost the small tensorial signal present within LSS probes. Unlike for scalars, the measurement of tensors in the primary CMB is not cosmic-variance limited (since the $B$-mode signal is, under null linear assumptions, zero), though pSZ can still add information by increasing the number of fundamental modes available. As described above, cross-correlations could be of use in making such a detection, since they do not suffer from many of the traditional systematic effects such as atmospheric dust absorption (since they contain only one power of the CMB), thus it is important to explore whether such statistics can be practically useful.

In the remainder of this work, we consider whether future surveys are capable of measuring the pSZ-shear cross-correlation. Such a detection could place further constraints on novel observables, be it stronger bounds on the ISW effect, the strongly-unequal-time SW effect, or tensor modes. After laying out our conventions in \S\ref{sec: conventions}, we will present the contributions to galaxy shear, galaxy density, and the pSZ effect from scalars and tensors in \S\ref{sec: signal} and from noise in \S\ref{sec: noise}. Our main results are forecasts on the detectability of the cross-spectra themselves and various physical components, which we present in \S\ref{sec: detectability}. In \S\ref{sec:cosmotests} we describe the novel properties illustrated in Fig.~\ref{fig: cartoon} in the context of a toy model before concluding in \S\ref{sec: discussion}. Appendix \ref{appen: transfer} lists the transfer functions used in this work, whilst Appendix \ref{appen: ksz} presents a brief forecast of the kSZ auto- and cross-correlations. All calculations are made publicly available at \href{github.com/oliverphilcox/pSZ-cross-Shear}{GitHub.com/OliverPhilcox/pSZ-cross-Shear}. 

\vskip 4pt

\section{Conventions}\label{sec: conventions}
We briefly present the various conventions for scalar and tensor perturbations used in this work, as well as for cosmic shear. Note that conventions differ between works, e.g., our results appear to differ from those of \citep{Deutsch:2017cja} and \citep{2012PhRvD..86h3513S} until notational variations are taken into account.

\subsection{Scalar Modes}\label{subsec: scalar-conventions}
We primarily work with the Newtonian potential, $\Psi$, which enters the FLRW metric in the standard fashion (in the conformal Newtonian gauge, with $c=1$):
\beq
    ds^2=a^2(\eta)\left[-(1+2\Psi)d\eta^2+(1-2\Psi)dx^idx_i\right],
\eeq
assuming the stress-free condition. The statistics of $\Psi$ are described by its power spectrum:
\beq\label{eq: Pphi-def}
    \av{\Psi(\vk,\chi)\Psi^*(\vk',\chi')} = \delD{\vk-\vk'}P_\Psi(k,\chi,\chi'),
\eeq
where $P_\Psi$ can be written in terms of the dimensionless spectrum $\P_\Psi$ via $P_\Psi(k,\chi,\chi')=D_{\Psi}(\eta_0-\chi)D_{\Psi}(\eta_0-\chi')(2\pi^2/k^3)\P_\Psi(k)$, where $D_{\Psi}$ is the potential growth function, $\eta_0$ is the conformal time today, and we absorb the necessary transfer functions into $\P_\Psi(k)$. Explicit forms for the potential growth function (on super- and sub-horizon scales) can be found in \citep{Erickcek:2008jp,Zhang:2015uta}, and we note that $D_\Psi(a)\to (9/10)D_{+}(a)/a$ in the subhorizon limit, for the  usual growth function $D_+(a)$. Practically, $P_\Psi$ can be obtained from the matter power spectrum computed by CLASS, rescaling by the ratio of potential and density growth factors.

We will also require the velocity power spectra. On sufficiently large scales, this is given by~\citep{Erickcek:2008jp}
\beq
    \mathbf{v}(\vr,\chi)=-D_v(\chi)/D_\Psi(\chi)\nabla\Psi(\vk,\chi)
\eeq
utilizing the velocity growth factor
\beq
    D_v(\chi) = \frac{2a^2H(a)}{H_0^2\Omega_m}\frac{y(\chi)}{4+3y(\chi)}\left[D_\Psi(\chi)+\frac{dD_\Psi(\chi)}{d\log a}\right],
\eeq
where $y(\chi) = a(\chi)/a_{\rm eq}$.

\subsection{Tensor Modes}\label{subsec: tensor-conventions}
We define the transverse-traceless tensor metric perturbation, $h_{ij}$, via
\beq
    ds^2=a^2(\eta)\left[-d\eta^2+(\delta_{ij}^{\rm K}+h_{ij})dx^idx^j\right].
\eeq
This is often written in terms of $+,\times$ states via
\beq
    h_{ij}(\vr,\eta) =  \tilde h_+(\vr,\eta) e_{ij}^{+}(\hat{\bv r})+ \tilde h_\times(\vr,\eta) e_{ij}^\times(\hat{\bv r}).
\eeq
In this work, we primarily expand in helicity states:
\beq\label{eq: tensor-helicity}
    h_{ij}(\vr,\chi) = \int_{\vk}e^{i\vk\cdot\vr}\sum_{\lambda\in\pm}h_\lambda(\vk,\chi) e^{(\lambda)}_{ij}(\hk),
\eeq
where $e_{ij}^{(\lambda)}e^{ij}_{(-\lambda)}=1$, $h_\lambda = e^{ij}_{(-\lambda)}h_{ij}$, and we notate $\int_{\vk}\equiv \int d\vk/(2\pi)^3$. Here, $e^{(\lambda)}_{ij}=e^{(\lambda)}_ie^{(\lambda)}_j$ with $\bv e^{(\lambda)}=(\bv e_1\mp i\lambda\bv e_2)/\sqrt{2}$ where $\{\hk,\bv e_1, \bv e_2\}$ form an orthonormal set. These are related to the $\tilde h_{+,\times}$ basis by $h_{\pm} = (\tilde h_+\mp i\tilde h_\times)/2$.

The statistics of $h$ are specified by
\beq
    \av{h_\lambda(\vk,\chi)h_\lambda^*(\vk',\chi')} = \delD{\vk-\vk'}P_{h_\lambda}(k,\chi,\chi'),
\eeq
with the total power spectrum $P_h=\left[P_{h_+}+P_{h_-}\right]$, and chiral spectrum $\Delta_hP_h=\left[P_{h_+}-P_{h_-}\right]$.\footnote{In the notation of \citep{Deutsch:2018umo}, $P_{h_+}=P_L/2$, $P_{h_-}=P_R/2$, $\Delta_c=-\Delta_h$ and $P_h^{\rm this \,work}=P_h^{\rm former}/2$.} This is related to the primordial spectrum $\P_h(k)$ via
\beq\label{eq: P-h}
    P_h(k,\chi,\chi') = \frac{2\pi^2}{k^3}D_{\rm T}(k,\eta_0-\chi)D_{\rm T}(k,\eta_0-\chi')\P_h(k),
\eeq
where $D_{\rm T}(k,\eta)\approx 3j_1(k\eta)/(k\eta)$ is the tensor transfer function, assuming matter domination. The tensor spectrum is usually parametrized as
\beq
    \P_h(k) = \Delta_{\rm T}^2\left(\frac{k}{k_\ast}\right)^{n_{\rm T}},
\eeq
where $n_{\rm T}\approx-r/8$ is the spectral index and $\Delta_{\rm T}^2=r\Delta_{\zeta}^2$ is the amplitude, for curvature perturbation $\zeta$, and characteristic scale $k_0$.

\subsection{Shear}\label{subsec: shear-conventions}
We define the components of the full-sky shear as ${}_{\pm 2}\gamma(\hn)=m_{\mp}^im_{\mp}^j\gamma_{ij}(\hn)$, where the basis vectors are
\beq
    \bv m_{\pm}=\frac{1}{\sqrt 2}\begin{pmatrix}\cos\theta\cos\varphi\pm i\sin\varphi\\ \cos\theta\sin\varphi\mp i\cos\varphi\\-\sin\theta\end{pmatrix}.
\eeq
The spherical harmonic coefficients are defined via
\beq
    {}_{+2}\gamma_{\ell m} = \sqrt{\frac{(\ell-2)!}{(\ell+2)!}}\int d\hn\,Y^*_{\ell m}(\hn)\bar\edth^2{}_{+2}\gamma(\hn),\qquad {}_{-2}\gamma_{\ell m} = \sqrt{\frac{(\ell-2)!}{(\ell+2)!}}\int d\hn\,Y^*_{\ell m}(\hn)\edth^2{}_{-2}\gamma(\hn),
\eeq
where $\edth$ and $\bar\edth$ are the usual spin-raising and spin-lowering operators \citep[e.g.,][]{2005PhRvD..72b3516C}. These are related to the $E$- and $B$-modes via ${}_{\pm 2}\gamma_{\ell m}=\gamma^E_{\ell m}\pm i\gamma^B_{\ell m}$. The corresponding power spectra are
\beq
    C_\ell^{\gamma^X\gamma^Y} = \frac{1}{2\ell+1}\sum_{m=-\ell}^\ell \av{\gamma^X_{\ell m}\gamma^{Y*}_{\ell m}},
\eeq
for $X\in\{E,B\}$, which is parity even (odd) if $X=Y$ ($X\neq Y$).

\section{Signal Modeling}\label{sec: signal}

In this section, we describe how to compute the signal auto- and cross-spectra for galaxy density, galaxy shear, and the remote quadrupole field, considering both scalar and tensor sources.

\vskip 8 pt
\paragraph{Galaxy Density}
The basic observable in a galaxy redshift survey is the galaxy overdensity. The overdensity in a shell at fixed comoving radial distance is $\delta_g(\chi\hn) = [n(\chi\hn)-n(\chi)]/n(\chi)$, for observed field $n(\chi\hn)$, and depends principally on the scalar potential $\Psi$. At linear order, we have the usual relation 
\beq
    \delta_g(\chi\hn) = b_g(\chi)\int_{\vk}e^{i\vk\cdot\chi\hn}\delta_m(\vk,\chi) =  -\frac{2a(\chi)}{3H_0^2\Omega_m}b_g(\chi)\int_{\vk}e^{i\vk\cdot\chi\hn}k^2\Psi(\vk,\chi),
\eeq
where $\delta_m(\vk,\chi)$ is the Fourier-space matter density, related to $\Psi$ via the Poisson equation, and we neglect relativistic effects. In this paper, we will consider photometric galaxy surveys where the galaxy density is measured in redshift bins labeled $a = 1, 2, \ldots N_{\rm bin}$ and given by
\beq\label{eq:binnedgal}
    \delta_{g,a} (\hn) = \int_0^\infty d\chi\,n_a(\chi) \delta_g(\chi\hn),
\eeq
where $n_a(\chi)\propto n(\chi)W_a(\chi)$ is the normalized source density in bin $a$, as before.\footnote{\resub{In the limit of $N_{\rm bin}\to\infty$ this is analogous to a three-dimensional spectroscopic sample, except without redshift effects.}} This uses the true source density $n(\chi)$ and a user-defined weighting function $W_a(\chi)$. Note that here and everywhere below we neglect redshift space distortions, magnification, and relativistic projection effects. There are only scalar contributions to this observable.

\vskip 8 pt
\paragraph{Galaxy Shear}
The shape distortions of galaxies are usually expressed using the shear tensor $\gamma_{ij}$ (neglecting higher-order moments such as flexion \citep[e.g.,][]{2006MNRAS.365..414B}). Roughly speaking, this is a measurement of a galaxy's ellipticity, and is usually projected onto the two-sphere by binning in redshift, \textit{i.e.} \
\beq\label{eq: z-integration}
    \gamma_{ij,a}(\hn) = \int_0^\infty d\chi\,n_a(\chi)\gamma_{ij}(\chi\hn),
\eeq
where $\chi\hn$ is the three-dimensional galaxy position at comoving distance $\chi$, and $n_a(\chi)\propto n(\chi)W_a(\chi)$ is the normalized source density in bin $a$, for source density $n(\chi)$, as before. Following projection, the shear tensor is a spin-two field and can be expressed in ${}_{\pm2}\gamma$ components or $E$- and $B$-modes (cf.\,\S\ref{subsec: shear-conventions}). 

The Newtonian potential $\Psi$ sources two contributions to galaxy shear: intrinsic alignments and weak lensing. For a source galaxy at redshift $\chi$, the spin-$\pm2$ shear components are given by \citep[e.g.,][]{2005PhRvD..72b3516C,2001PhR...340..291B,2004PhRvD..70f3526H,2017JCAP...05..014L}:
\beq\label{eq:binnedscalarshear}
    {}_{\pm 2}\gamma_{{\rm S}, a} (\hn)  &=& \int_0^\infty d\chi\,n_a(\chi)\ \left[ -b_{\rm S}(\chi)m_\mp^im_\mp^j\Psi_{,ij}(\chi\hn,\chi) + 2\int_0^\chi\frac{d\chi'}{\chi'}\frac{\chi-\chi'}{\chi}m_\mp^im_\mp^j\Psi_{,ij}(\chi'\hn,\chi') \right],
\eeq
where $m_\pm$ are the basis components given in \S\ref{subsec: shear-conventions}. The first term is the intrinsic alignment contribution (arising from galaxies preferentially aligning with a local tidal field), and involves the intrinsic alignment bias $b_{\rm S}(\chi)\equiv(2/3)C^{\rm S}_{1}\rho_{cr0}H_0^{-2}$ (in the notation of \citep{2011JCAP...05..010B,2012PhRvD..86h3513S}), with $\rho_{cr0}=3H_0^2/(8\pi G)$, and $C_1^{\rm S}\rho_{cr0}\sim 0.1$. The second term is from weak lensing, and involves the integrated scalar perturbation along the photon's worldline from the source galaxy to the observer. 

Following \citep{2012PhRvD..86h3513S}, tensor modes source the following contributions to galaxy shear:
\beq\label{eq:binnedtensorshear}
    {}_{\pm 2}\gamma_{{\rm T}, a} (\hn) &=&  \int_0^\infty d\chi\,n_a(\chi)\ \left[ -\frac{1}{2}h_{\pm}(\bv 0,0)-\frac{1}{2}\left[1-b_{\rm T}(\chi)a^{-2}(\chi)(\partial_\eta^2+aH\partial_\eta)\right]h_{\pm}(\chi\hn,\chi) \right. \\ \nonumber
    &&\left. -\int_0^\chi d\chi'\,\left[\frac{\chi-\chi'}{2}\frac{\chi'}{2\chi}m_{\mp}^im_{\mp}^jh_{kl,ij}\hat n^k\hat n^l+\left(1-\frac{2\chi'}{\chi}\right)\hat n^lm_\mp^km^i_\mp h_{kl,i}-\frac{1}{\chi}h_{\pm} (\chi'\hn,\chi') \right] \right]
\eeq
for $h_{\pm}\equiv m_{\mp}^im_{\mp}^jh_{ij}$, where our tensor conventions are specified in \S\ref{subsec: tensor-conventions}. The first and second terms on the first line correspond to observer and source distortions (frame of reference effects), the third term is from intrinsic alignments, and the second line gives contributions integrated along the line of sight from the source to the observer, \textit{i.e.}\ weak lensing effects. The coefficient $b_{\rm T}(\chi)\equiv (2/3)C_1^{\rm T}\rho_{cr0}H_0^{-2}$ specifies the strength of the alignment effect with $C_1^{\rm S}\sim C_1^{\rm T}$ expected in practice (though see \citep{2014PhRvD..89h3507S,2020JCAP...07..005B} for further discussion of this).

\vskip 8 pt
\paragraph{Remote Quadrupole}
The polarized Sunyaev Zel'dovich (pSZ) effect sources polarization anisotropies through the Thomson scattering of CMB photons from the locally observed CMB quadrupole seen by free electrons in the post-reionization Universe (the remote quadrupole field). Given a tracer of the optical depth, such as a galaxy redshift survey, and high-resolution measurements of the CMB polarization, it is possible to reconstruct the remote quadrupole field using a quadratic estimator as described in \citep{Alizadeh:2012vy,Deutsch:2017ybc,Deutsch:2018umo}. In essence, this estimator probes the combination $\delta_{g,a} (Q\pm iU)$, thus its auto-spectra is really a four-point function of the form $\av{\delta_{g,a}^2(Q\pm iU)^2}$. The remote quadrupole field at some position $\chi\hn$ is defined by
\beq\label{eq: remote-q}
    \Theta_{2m} (\chi\hn) = \int d\hn\,\ \Theta(\chi\hn,\hn') \ Y_{2m}^*(\hn'),
\eeq
where $\hn'$ is the emission angle and $\Theta$ is the temperature perturbation. There are contributions to $\Theta_{2m}$ from both scalar and tensor modes via the usual Sachs Wolfe, Integrated Sachs Wolfe, and Doppler components. The quantity reconstructed in pSZ tomography is a projection of the remote quadrupole field onto our line of sight, and integrated over radial bins with the same weighting as the galaxy density:
\beq
q^{\pm}_a (\hn) = \int_0^\infty d\chi\,n_a(\chi) \ \sum_{m=-2}^2 \Theta_{2m} (\chi\hn) \ {}_{\mp 2} Y_{2m}(\hn),
\eeq
where ${}_{\mp 2} Y_{2m}(\hn)$ are spin-weighted spherical harmonic. As for shear, we can form $E$ and $B$ modes in harmonic space as $q^{\pm}_{\ell m, a} = q^{E}_{\ell m, a} \pm i q^{B}_{\ell m, a}$. Scalar perturbations source only an $E$-mode, while tensors source both $E$ and $B$ modes. 

\vskip 8 pt
\paragraph{Spectra}
Having defined each of the observables above, can write a general element of the signal covariance in terms of an integral of transfer functions convolved with primordial spectra. For scalars, we have:
\beq\label{eq:scalar_cov}
    \left.C_{\ell, ab}^{X Y}\right|_{\rm S} &=& 4\pi\int_0^\infty d\log k\,\ \Delta_{\ell, a}^{X,\rm{S}}(k,\chi) \Delta_{\ell, b}^{Y,\rm{S}*}(k,\chi') \P_\Psi(k),
\eeq
where $X,Y \in \{\delta_{g}, \gamma^E, q^{E}\}$. The explicit form of the transfer functions $\Delta_{\ell, a}^{X,\rm{S}}(k,\chi)$ is presented in Appendix~\ref{appen: transfer}. For tensors, we find a similar form:
\beq\label{eq:tensor_cov}
    \left.C_{\ell, ab}^{X Y} \right|_{\rm T} &=& 4\pi\int_0^\infty d\log k\,\ \Delta_{\ell, a}^{X,\rm{T}}(k,\chi) \Delta_{\ell, b}^{Y,\rm{T}*}(k,\chi') \P_h(k)\times\left[\delta_{\rm K}^{XY}+(1-\delta_{\rm K}^{XY})\Delta_h\right],
\eeq
where $\delta_{\rm K}^{XY}$ is the Kronecker delta, the last term encodes chirality, $X,Y \in \{\gamma^E, q^{E}, \gamma^B, q^{B}\}$, and we assume $\ell\geq2$. The explicit form of the tensor transfer functions $\Delta_{\ell, a}^{X,\rm{T}}(k,\chi)$ is again given in Appendix~\ref{appen: transfer}. 

\section{Noise Modeling}\label{sec: noise}

For the forecasts presented below, we must make some assumptions about the hypothetical galaxy survey and CMB experiment used to measure the galaxy density, shear, and remote quadrupole field. The galaxy density and shear are limited by the mean number density of objects in the survey, $\bar n$; for the galaxy density, this sources noise spectra of the form
\beq
    \left.C_{\ell, ab}^{\delta_{g} \delta_{g}}\right|_{\rm noise} = \frac{1}{\bar n_a}\delta^{\rm K}_{ab},
\eeq
where $\bar n_a^{-1} = \int_0^\infty d\chi\,n(\chi)W_a^2(\chi)/\left[\int_0^\infty d\chi\,n(\chi)W_a(\chi)\right]^2$. For shear, the noise spectra are given by
\beq
    \left.C_{\ell,ab}^{\gamma^X\gamma^Y}\right|_{\rm noise} = \frac{\sigma^2_\gamma}{\bar n_a}\delta^{\rm K}_{ab}\delta_{\rm K}^{XY},
\eeq
with $X,Y \in \{E,B\}$. 

The noise on the reconstructed remote quadrupole field is somewhat more complex, as it is dependent on the estimator, galaxy survey, CMB experiment, and signal spectra. The quadratic estimator for the remote dipole is of the form
\beq
    \widehat{q}^{X}_{\ell m,a} = \sum_{\ell_1m_1\ell_2m_2}\left(W_{\ell m\ell_1m_1\ell_2m_2}^{X,E}a_{\ell_1m_1}^E+W_{\ell m\ell_1m_1\ell_2m_2}^{X,B}a_{\ell_1m_1}^B\right)\Delta\tau_{\ell_2 m_2,a},
\eeq
for $X\in\{E,B\}$ (using the decomposition of \S\ref{subsec: shear-conventions}), where $a_{\ell m}^{E,B}$ are the CMB $E$- and $B$-modes, $\Delta\tau_a$ is the optical depth in bin $a$, and $W$ are some weight matrices whose form can be found in \citep{Deutsch:2017ybc}. From \citep{Deutsch:2017cja}, the estimator noise is given by
\beq\label{eq: noise-pSZ}
    \frac{1}{\left.C_{\ell,ab}^{q^Eq^E}\right|_{\rm noise}} &=& \frac{\delta_{\rm K}^{ab}}{2\ell+1}\sum_{\ell_1\ell_2}\frac{\Gamma_{\ell\ell_1\ell_2,a}^{\rm pSZ}\Gamma_{\ell\ell_1\ell_2,b}^{\rm pSZ}}{\left(|\alpha_{\ell\ell_1\ell_2}|^2C_{\ell_1}^{EE}+|\gamma_{\ell\ell_1\ell_2}|^2C_{\ell_1}^{BB}\right)C_{\ell_2,ab}^{\delta_g\delta_g}}\\\nonumber
    \frac{1}{\left.C_{\ell,ab}^{q^Bq^B}\right|_{\rm noise}} &=& \frac{\delta_{\rm K}^{ab}}{2\ell+1}\sum_{\ell_1\ell_2}\frac{\Gamma_{\ell\ell_1\ell_2,a}^{\rm pSZ}\Gamma_{\ell\ell_1\ell_2,b}^{\rm pSZ}}{\left(|\gamma_{\ell\ell_1\ell_2}|^2C_{\ell_1}^{EE}+|\alpha_{\ell\ell_1\ell_2}|^2C_{\ell_1}^{BB}\right)C_{\ell_2,ab}^{\delta_g\delta_g}},
\eeq
with vanishing $\left.C_{\ell, ab}^{q^Eq^B}\right|_{\rm noise}$ due to parity conservation. In the above, $|\alpha|^2$ ($|\gamma|^2$) is one if $\ell+\ell_1+\ell_2$ is even (odd) and zero else, and the CMB spectra $C_{\ell}^{EE,BB}$ include lensing and noise. Here, the weighting function is given by
\beq
    \Gamma_{\ell\ell_1\ell_2, a}^{\rm pSZ} = -\frac{\sqrt{6}}{10}\sqrt{\frac{(2\ell+1)(2\ell_1+1)(2\ell_2+1)}{4\pi}}\tj{\ell}{\ell_1}{\ell_2}{2}{-2}{0}C_{\ell_2, ab}^{\Delta \tau \delta_{g}},
\eeq
where the quantity in parentheses is a Wigner $3j$ symbol. The galaxy-galaxy and galaxy-optical depth spectra used in the noise computation, assuming that ionized gas traces dark matter on all scales,\footnote{This is related to the so-called optical depth bias. Because the pSZ signal is dependent on the optical depth, and we reconstruct it with an imperfect tracer (galaxies in this case), there is modelling uncertainty implicit in the estimator that can bias the reconstructed remote quadrupole field. We do not incorporate an analysis of the optical depth degeneracy in this work.} are given by:
\beq\label{eq: delta-g-tau-Cl}
    C_{\ell, ab}^{\delta_g\delta_g} &=& 4\pi\int_0^\infty d\log k\,\left[\int_0^\infty d\chi\,n_a(\chi) \ b_g(\chi)D_+(\eta_0-\chi)j_\ell(k\chi)\right]\left[\int_0^\infty d\chi' \,n_b(\chi') \ b_g(\chi')D_+(\eta_0-\chi')j_\ell(k\chi')\right]\P_{\delta}(k) \nonumber\\ 
    &+& \frac{1}{\bar n_a}\delta^{\rm K}_{ab}\\ 
    C_{\ell, ab}^{\Delta\tau\delta_g} &=& 4\pi\int d\log k\,\left[\int_0^\infty d\chi \ \sigma_Ta(\chi)W_a(\chi)n_e(\chi)D_+(\eta_0-\chi)j_\ell(k\chi)\right]\left[\int_0^\infty d\chi'\,n_b(\chi') b_g(\chi')D_+(\eta_0-\chi')j_\ell(k\chi')\right]\P_{\delta}(k) \nonumber
\eeq
where we introduce the matter power spectrum $\P_\delta(k)\equiv (k^3/2\pi^2)P_\delta(k,z=0)$ with growth factor $D_+$, electron density $n_e(\chi)$, and the Thomson cross-section $\sigma_T$. 
Since we require these spectra at large $\ell$, they are computed with the Limber approximation (including non-linear effects only through $\P_\delta$).

\section{Detectability Forecasts}\label{sec: detectability}
We now turn to the question of whether the above effects are measurable in practice. To ascertain this, we will consider a simple forecast appropriate for next-generation observatories, aiming to measure both the signal-to-noise of the various signals and the detectability of cosmological parameters such as the tensor-mode amplitude.

For these forecasts, we assume the following:
\begin{itemize}
    \item \textbf{Galaxy Sample}: VRO-like (LSST Gold sample), with $n(z) \propto z^2\exp(-z/z_0)$ for $z_0 = 0.3$ with a source density of $40\,\mathrm{arcmin}^{-2}$, ignoring photometric redshift errors \citep{LSSTScience:2009jmu}. We assume linear bias \resub{$b(z)=0.95/D_+(z)$} for growth factor $D_+(z)$ and a sky fraction of $f_{\rm sky} = 0.36$.\footnote{As noted in \citep{Ferraro:2022twg}, a deeper sample of galaxies can be obtained from VRO by including drop-outs with a higher magnitude limit. Whilst this would aid the detectabilities considered herein, we do not include it, since the additional high-redshift galaxies will not be measured at sufficiently high resolution to enable shear measurement. \resub{Furthermore, we note that the forms of $n(z)$ and $b(z)$ are uncertain at high redshift. Since the high-$z$ data add little constraining power (cf.\,\S\ref{subsec: scalar-detect}), this is unlikely to significantly affect our forecast.}}
    \item \textbf{Binning}: Six bins with $z\in[0.1,6]$, each a top-hat in comoving distance with $\Delta\chi \approx 1600\,\mathrm{Mpc}$. The inverse galaxy density varies from $6\,\mathrm{arcmin}^{-2}$ to $0.0002\,\mathrm{arcmin}^{-2}$ from low to high redshift. We include all necessary cross-covariances in our forecasting, and will discuss the dependence on $z_{\rm max}$ and the number of bins below.
    \item \textbf{CMB}: Gaussian instrumental noise and beam, taking the standard form $N_\ell^{E,B} = \Delta_P^2\,\mathrm{exp}\left[\ell(\ell+1)\theta_{\rm FWHM}/8\log 2\right]$, for noise $\Delta_P = 1\,\mu\mathrm{K}\text{-}\mathrm{arcmin}$ and beam width $\theta_{\rm FWHM} = 1\,\mathrm{arcmin}$, as appropriate for CMB-S4 \citep{Abazajian:2019tiv}. We will also consider a higher resolution sample with $\Delta_P = 0.5\,\mu\mathrm{K}\text{-}\mathrm{arcmin}$ noise, a $0.25\,\mathrm{arcmin}$ beam and a source density of $100\,\mathrm{arcmin}^{-2}$ over $f_{\rm sky}=0.5$, similar to CMB-HD \citep{Sehgal:2019ewc}.\footnote{Note that we neglect any leakage between temperature and polarization in the CMB experiments; this could potentially lead to a percent-level bias in the high-$\ell$ polarization spectra, which would have severe implications for pSZ detectability.}
    \item \textbf{Cosmological Parameters}: For $\Lambda$CDM: $\{h = 0.7, A_s = 1.95\times 10^{-9}, n_s = 0.96, \Omega_b = 0.049, \Omega_m = 0.3\}$. For tensors, we set $r_{\rm fid} = 1$ for illustration, with $k_{0} = 0.05\,\mathrm{Mpc}$ and $n_t = -r/8$. Where relevant, we assume maximal chirality, \textit{i.e.}\ $\Delta_h = 1$.
\end{itemize}

All power spectra are computed in \textsc{python} via explicit integration of the kernels given above against scalar and tensor power spectra, given by \textsc{class}, using all $\ell$ in the range $[2,100]$ (noting that the signal-to-noise falls quickly with $\ell$). For the pSZ noise spectra, we compute the lensed CMB spectra up to $\ell=10^4$ using \textsc{class}, with the optical depth spectra computed assuming that $n_e(\chi) = (7/8)n_b(\chi)$ (appropriate for a hydrogen fraction of $3/4$), and assume full ionization with the electron inhomogeneities tracing those of matter, ignoring optical depth degeneracies. To evaluate \eqref{eq: noise-pSZ} we perform a direct sum over $\ell_1,\ell_2$ for $\ell_i\in[1,10000]$, with the relevant Wigner $3j$ symbols precomputed using recurrence relations,\footnote{We use the implementation of \href{github.com/xzackli/WignerFamilies.jl}{github.com/xzackli/WignerFamilies.jl}.} and the high-$\ell$ galaxy and optical depth spectra computed using the Limber approximation \citep[e.g.,][]{2017JCAP...05..014L}. \textsc{Jupyter} notebooks containing all the analysis code (and a number of associated plots) can be found at \href{github.com/oliverphilcox/pSZ-cross-Shear}{GitHub.com/OliverPhilcox/pSZ-cross-Shear}.

In the Gaussian limit, the various signals (including noise contribtions) have covariance:
\beq\label{eq: signal-cov}
    \mathrm{Cov}\left(C_{\ell,ab}^{XY},C_{\ell',cd}^{ZW}\right) = \frac{2\delta^K_{\ell\ell'}}{(2\ell+1)f_{\rm sky}}\left[C_{\ell,ac}^{XZ}C_{\ell,bd}^{YW}+C_{\ell,ad}^{XW}C_{\ell,bc}^{YZ}\right],
\eeq
which is diagonal in $\ell$. When considering the detectability of pSZ signals, only noise and scalar shear spectra appear on the RHS of \eqref{eq: signal-cov}, but when considering tensors, we include also scalar pSZ contributions as an effective noise term. The Fisher matrix takes the standard form \citep[e.g.,][]{1997ApJ...480...22T};
\beq\label{eq: fisher}
    F_{\alpha\beta} = \sum_\ell\frac{d\mathbf{D}_\ell}{dp_\alpha}\mathbf{C}_\ell^{-1}\frac{d\mathbf{D}_\ell}{dp_\beta},
\eeq
for some set of parameters $\{p_\alpha\}$, where the data-vector, $\mathbf{D}_\ell$, and covariance $\mathbf{C}_\ell$, contain all non-trivial auto- and cross-spectra. Using six tomographic bins, we find a total of 42 (156) parity-even and 36 (144) parity-odd spectra for pSZ or shear (pSZ and shear). Under null assumptions, parity-odd and parity-even spectra are uncorrelated, thus we may compute their contributions to Fisher forecasts separately. Via the Cramer-Rao bound, the $1\sigma$ bound on $p_\alpha$ satisfies $\sigma^2_{p_\alpha}\geq \left(F^{-1}\right)_{\alpha\alpha}$.

\subsection{Numerical Results}
\begin{figure}
    \centering
    \includegraphics[width=\textwidth]{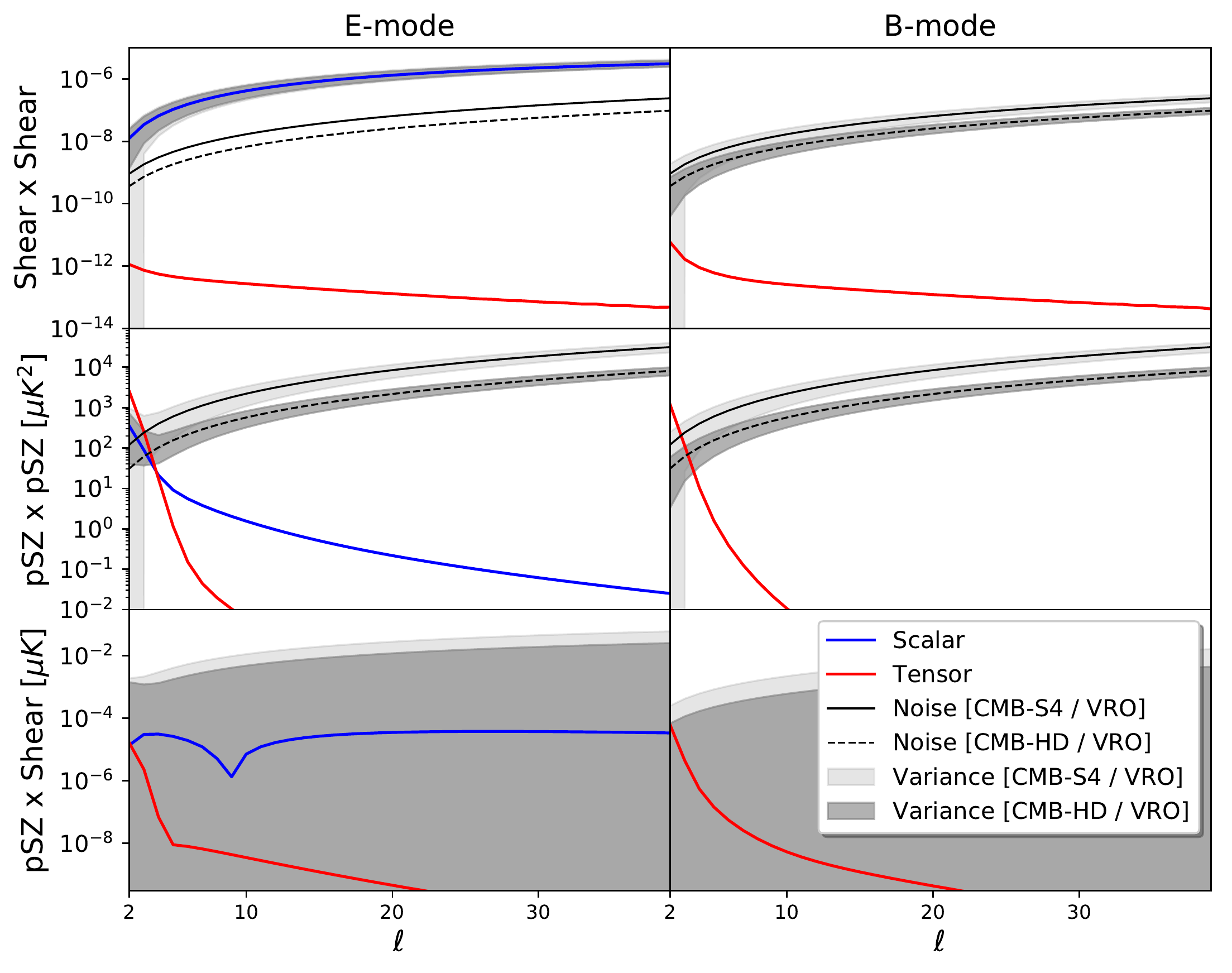}
    \caption{Contributions of scalar and tensor modes to the shear and pSZ angular power spectrum. We show a single redshift bin centered at $z=1$, and plot scalar (tensor) contributions to the spectra in blue (red, assuming $r=1$). The first, second, and third plots show the shear-auto, pSZ-auto and shear-pSZ spectra for $E$-modes (left) and $B$-modes (right), with all spectra multiplied by $\ell(\ell+1)/(2\pi)$. \resub{We additionally convert the CMB based measurements into micro-Kelvin units by multiplying by $T_{\rm CMB}$.} The black curves show noise contributions, and the gray regions give $1\sigma$ errors around the total tensor-free signal, relevant for both CMB-S4/VRO and CMB-HD/VRO experiments. Clearly, the pSZ-shear cross-spectra has much reduced signal-to-noise compared to the pSZ-pSZ spectra, though some detection of gravitational waves may be possible in the $B$-mode, particularly at larger $z$.}
    \label{fig: combined-EB}
\end{figure}

Fig.\,\ref{fig: combined-EB} displays the auto- and cross-spectra of pSZ and galaxy shear for a single redshift bin, separating out scalar, tensor, and noise components. As expected, the shear-shear spectra contain strong scalar contributions (which form the workhorse of many previous $S_8-\Omega_m$ analyses), but, as in \citep{2012PhRvD..86h3513S}, only very weak contributions from tensors. Even in the $B$-mode (which is not cosmic-variance limited), the signature of $r=1$ gravitational waves can be orders-of-magnitude below the noise floor of CMB-S4/VRO, and accessible only at the smallest $\ell$, where foreground and systematic effects are most important. At higher redshift, shear is of greater use, though upcoming photometric surveys are optimized only for the relatively local Universe.

For pSZ auto-spectra, the noise threshold remains a significant limitation, but, as seen in the middle panel of Fig.\,\ref{fig: combined-EB}, both scalars and tensors can be potentially detected on very large scales, matching the results of previous work \citep[e.g.,][]{Deutsch:2017cja,Deutsch:2017ybc}. In contrast to the shear auto-correlation, the tensor spectra is relatively evenly split between $E$- and $B$-modes; this occurs since the scalar $E$-mode contribution is weak, thus the pSZ noise limits both samples.

The cross-spectra paint a somewhat different picture. In this case, there is no experimental noise curve (since $\av{E} = \av{B} = 0$ in the CMB), but significant variance, even for futuristic experimental set-ups based on CMB-HD. That said, both scalar and tensor contributions are clearly non-zero, with the latter peeking above the cosmic variance in the large-scale $B$-mode. As discussed below, the various contributions are a strong function of redshift, but the trend of Fig.\,\ref{fig: combined-EB} is relatively general: both scalar and tensor cross-spectra exist but will be difficult to detect. This is quantified in the following sections.

\begin{figure}
    \centering
    \includegraphics[width=\textwidth]{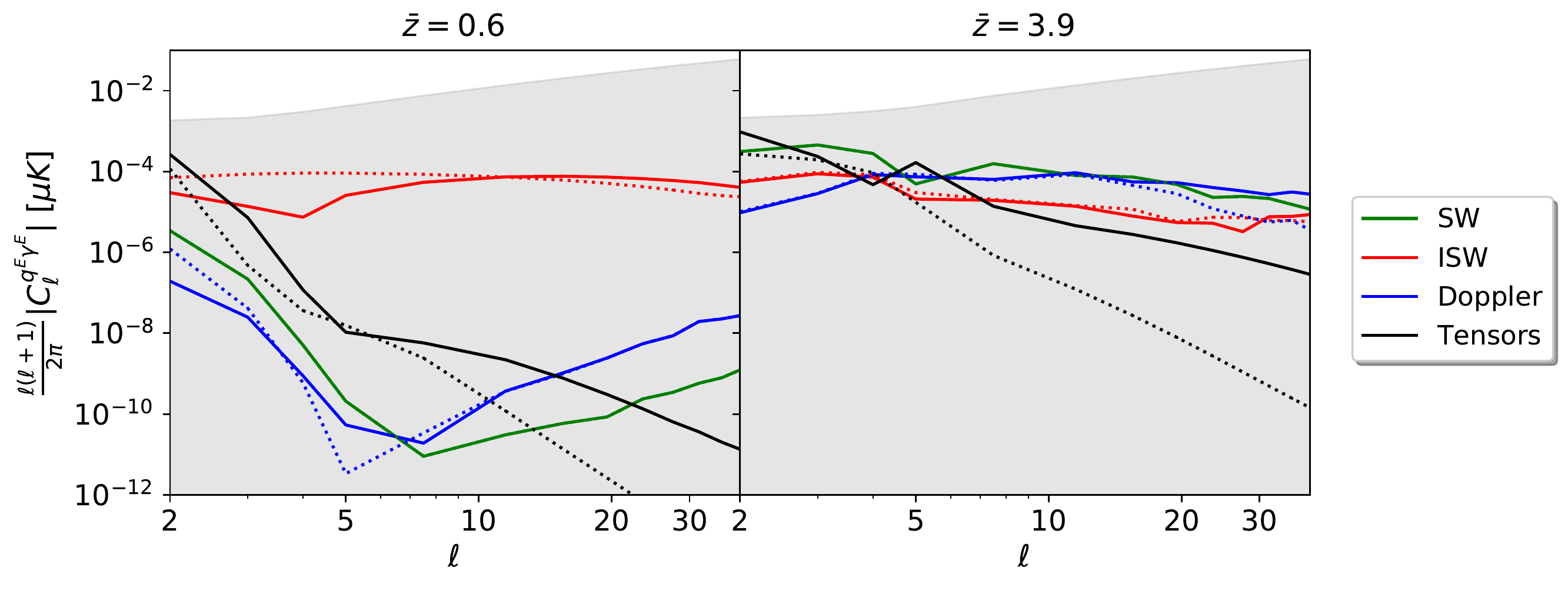}
    \caption{Contributions to the $E$-mode pSZ-shear cross-spectrum in two redshift bins, with centers indicated by the titles. We show scalar contributions from the Sachs-Wolfe (green, SW), integrated Sachs-Wolfe (red, ISW) and Doppler (blue) effect and $r=1$ tensors (black, as in Fig.\,\ref{fig: combined-EB}). The dotted lines show the same contributions but only including the effects of gravitational lensing (omitting intrinsic alignments). The shaded region shows the $1\sigma$ error for VRO / CMB-S4 as in Fig.\,\ref{fig: combined-EB}. We note that the ISW effect dominated the scalar signal at low-$z$, but there is significant contribution from the SW effect at high $z$. Furthermore, intrinsic alignments are seen to be a key contribution to the tensor signal in both bins, though the high-$z$ scalar signals are sourced almost entirely by lensing shear-modes.}
    \label{fig: qg-contributions}
\end{figure}

In Fig.\,\ref{fig: qg-contributions} we consider the cross-spectra in more detail, displaying results at both low- and high-redshift, and splitting the sample into the various contributions. At low redshifts, we see that the scalar contribution is dominated by the ISW effect, and contains power across a range of multipoles. This differs significantly from the pSZ auto-spectra, which is dominated by the SW effect and has power only at very low $\ell$. In principle, this suggests that the low-redshift shear-pSZ (or galaxy-pSZ) correlation could be a useful probe of the ISW effect. At higher redshifts, this is no longer the case; we find that the SW effect dominates over ISW, with the Doppler contribution being significantly suppressed regardless of redshift. This is as expected: the ISW effect occurs due to the time variation of gravitational potentials induced by dark energy, whose action is strongly suppressed for $z\gtrsim 2$. The signature of tensors appears similar to scalars: the most prominent signatures are observed at low-$\ell$, and, partly due to the greater impact of tensors on high-$z$ lensing, is most prominent at the largest redshifts. Finally, we consider the contribution of different lensing contributions: as shown in the figure, the scalar high-$z$ sample is dominated by the lensing correlations \citep[cf.\,][]{2012PhRvD..86h3513S}, whilst for scalars at low-$z$ and for tensors at all redshifts, intrinsic alignments are an important contributor to the signal, although we caution that they are accompanied by a poorly-understood bias parameter $b_{\rm T}(\chi)$.

\subsection{Detectability of Scalar pSZ}\label{subsec: scalar-detect}
To forecast the detection strength of pSZ we perform a Fisher forecast utilizing both the pSZ auto- and cross-spectra. For this purpose, we rescale the pSZ signal as $q^{E}\to (\alpha/\sqrt{2})q^{E}$, such that a Fisher forecast for $\alpha$ about $\alpha=1$ gives the desired signal-to-noise ratio (setting $\alpha=0$ in the covariance, \textit{i.e.}\ working under null assumptions). The following spectra contain scalar pSZ signatures:
\beq
    (\alpha^2/2)C_\ell^{q^Eq^E}\,,\,(\alpha/\sqrt 2)C_\ell^{q^E\gamma^E}\,,\,(\alpha/\sqrt 2)C_\ell^{q^Eg};
\eeq
these form the derivative vector in \eqref{eq: fisher} (summing over bins and multipoles). By considering only subcomponents of the pSZ spectra (cf.\,\S\ref{sec: signal}), we can additionally quantify the detection significance of physical signals such as the ISW effect.

\begin{figure}
    \centering
    \includegraphics[width=0.9\textwidth]{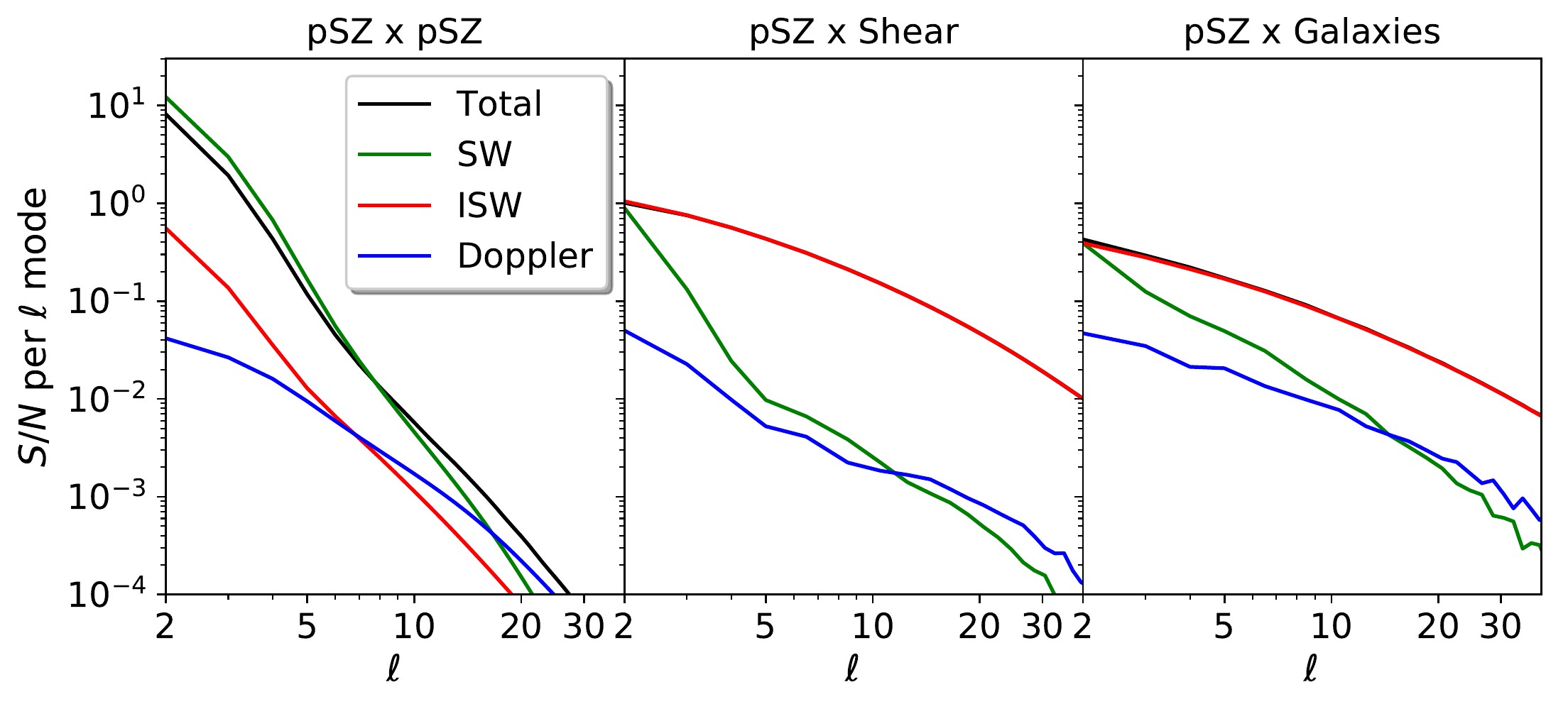}
    \caption{Signal-to-noise ($S/N$) ratio for the pSZ auto- and cross-spectra. We give the significance of a detection in each $\ell$ bin (equivalent to the $S/N$ per $\{\ell,m\}$ mode multiplied by $(2\ell+1)$); the total $S/N$ is the sum over all bins and given in Tab.\,\ref{tab: scalar-pSZ}. We show detection significances for the total pSZ effect (black) and split into the SW, ISW and Doppler subcomponents (green, red, and blue), always assuming the CMB-S4/VRO noise parameters. The overall signal is strongest in the auto-spectrum; however, this is more susceptible to systematic effects than cross-spectra. The auto-spectra are dominated by the SW signal, whilst cross-spectra are a probe instead of the ISW. Whilst all curves are a strong function of $\ell$, the fall-off is slower for the cross-spectra, indicating the utility of measuring smaller scales.}
    \label{fig: SN-scalar}
\end{figure}

\begin{table}
    \renewcommand{\arraystretch}{1.3}
    \caption{Signal-to-noise ratio of the scalar pSZ effect from auto- and cross-spectra with shear and galaxies. We split the signal into various components, and compute the signal-to-noise using the Fisher formalism of \eqref{eq: fisher}, assuming zero fiducial pSZ signal. Results are shown for two choices of CMB noise ($\Delta_P$, in $\mu\mathrm{K}\text{-}\mathrm{arcmin}$) and source density ($\bar n$, in $\mathrm{arcmin}^{-2}$). We find that future surveys will be able to detect the pSZ auto-spectrum at high significance (sourced by the SW effect), but cross-spectra with galaxies are practically unobservable, and those with shear are weak, though dominated by the ISW effect.\label{tab: scalar-pSZ}}
    \centering
    \begin{tabular}{l|cc|ccc|c}
    \hline\hline
    & $\Delta_P$ & $\bar{n}$ & SW & ISW & Doppler & Total\\\hline
    pSZ $\times$ pSZ & 1 & 40 & 12.59 & 0.58 & 0.05 & 8.40\\
    pSZ $\times$ Shear & 1 & 40 & 0.91 & 1.61 & 0.06 & 1.58\\
    pSZ $\times$ Galaxies & 1 & 40 & 0.42 & 0.61 & 0.07 & 0.65\\\hline
    pSZ $\times$ pSZ & 0.5 & 100 & 56.42 & 2.49 & 0.24 & 37.74 \\
    pSZ $\times$ Shear & 0.5 & 100 & 3.32 & 5.18 & 0.21 & 5.16\\
    pSZ $\times$ Galaxies & 0.5 & 100 & 1.07 & 1.33 & 0.18 & 1.44\\\hline\hline
    \end{tabular}
\end{table}

Our main results are given in Tab.\,\ref{tab: scalar-pSZ} and Fig.\,\ref{fig: SN-scalar}, both for the fiducial CMB-S4/VRO survey considered above, and a more futuristic survey based on CMB-HD/VRO. In each case, the total pSZ auto-spectrum can be robustly extracted (at $8.3\sigma$ and $37\sigma$ respectively), and is dominated by the SW effect; the other contributions are unmeasurable except for the ISW effect with futuristic noise levels. The signal-to-noise curves are a strong function of $\ell$: only the $\ell\lesssim5$ modes are recoverable by these techniques. 

The situation is more bleak for the cross-spectra. Combining pSZ and photometric galaxy density does not yield an observable signal for either choice of noise curves, and, further, the pSZ-shear correlation is small ($\approx1.6\sigma$) for CMB-S4/VRO noise levels. Further in the future however, we forecast a detection significance of $5.2\sigma$ for this cross-correlation using CMB-HD/VRO, which is dominated by the ISW effect. Noticeably, the decay of the signal-to-noise in cross-spectra is weaker than for auto-spectra; this indicates how more modes could be measured if the noise was particularly suppressed. Although the overall signal-to-noise is weak, it indicates how one, at least in principle, can extract the ISW effect from the usually-SW-dominated pSZ signal by utilizing cross-correlations.

It is important to ask whether these results depend on the redshift binning strategy adopted. To this end, we have performed an analogous Fisher forecast using 30 tomographic bins rather than $6$, each with a width of $\approx 300\Mpch$. Though such narrow bins are unlikely to be used in future optical surveys (due to photometric redshift uncertainties), they show how our results depend on the pSZ binning, and, for the galaxy cross-correlations, give an indication of how a spectroscopic survey would perform. In this case, we find very similar detection significances for all quantities, with an enhancement only at the $\lesssim 20\%$ level when using fine bins. In particular, the total pSZ-shear correlation can be detected at $1.8\sigma$ ($4.5\sigma$), whilst the pSZ-galaxy correlation is becomes $0.8\sigma$ ($1.8\sigma$) for CMB-S4/VRO (CMB-HD/VRO). We may similarly assess the dependence on the maximum survey redshift: we find only a small ($\lesssim10\%$) loss of signal-to-noise from reducing the redshift range to $[0.1,2]$ instead of $[0.1,6]$. This is due to the paucity of high-redshift objects in the fiducial sample. Altogether, the two tests indicate that six redshift bins are likely sufficient in practice, and that bins containing very few galaxies do not contribute significantly. \resub{Furthermore, we find that galaxy density is unlikely to be of practical use for measuring cross-correlations with pSZ, even if one uses a spectroscopic sample. Though the galaxy density field has lower noise than the shear observable (for photometric samples; far fewer sources are typically observed in spectrosopic analyses), the signal-to-noise of the cross-spectrum is dominated by modes in the linear regime (Fig.\,\ref{fig: SN-scalar}), which are instead cosmic-variance dominated. The difference in signal-to-noise indicates that the redshift kernel intrinsic to the remote quadrupole has better overlap with that from cosmic shear than galaxy density, and the limited impact of binning indicates that the signal is smooth in redshift and dominated by modes perpendicular to the line-of-sight.}
%, (c) utilizing a spectroscopic galaxy sample instead of a photometric one is unlikely to yield particular improvements (especially given the much reduced source density). 

\subsection{Detectability of Parity-Even Tensors}
We now turn to gravitational waves, considering the possible bounds upcoming and futuristic surveys can place on the tensor-to-scalar ratio $r$. Whilst one could also probe the tensor tilt, $n_{\rm T}$, this requires first measuring non-zero $r$, thus we will neglect it here (though see \citep{Deutsch:2018umo} for discussion of constraints from the pSZ auto-spectra). Assuming tensors are parity-conserving, gravitational wave signatures appear in the following spectra, all proportional to $r$:
\beq
    C_\ell^{q^Eq^E}\,,\,C_\ell^{q^Bq^B}\,,\,C_\ell^{\gamma^E\gamma^E}\,,\,C_\ell^{q^E\gamma^E}\,,\,C_\ell^{q^B\gamma^B}.
\eeq
Importantly, this involves $B$-modes, which via parity-conservation, do not contain contributions from scalars at leading-order, and thus provide a cleaner dataset within which to search for tensors. Additionally, there is no signal in the galaxy-pSZ cross-spectra (in linear theory), since the galaxy distribution is a scalar quantity.

\begin{figure}
    \centering
    \includegraphics[width=0.9\textwidth]{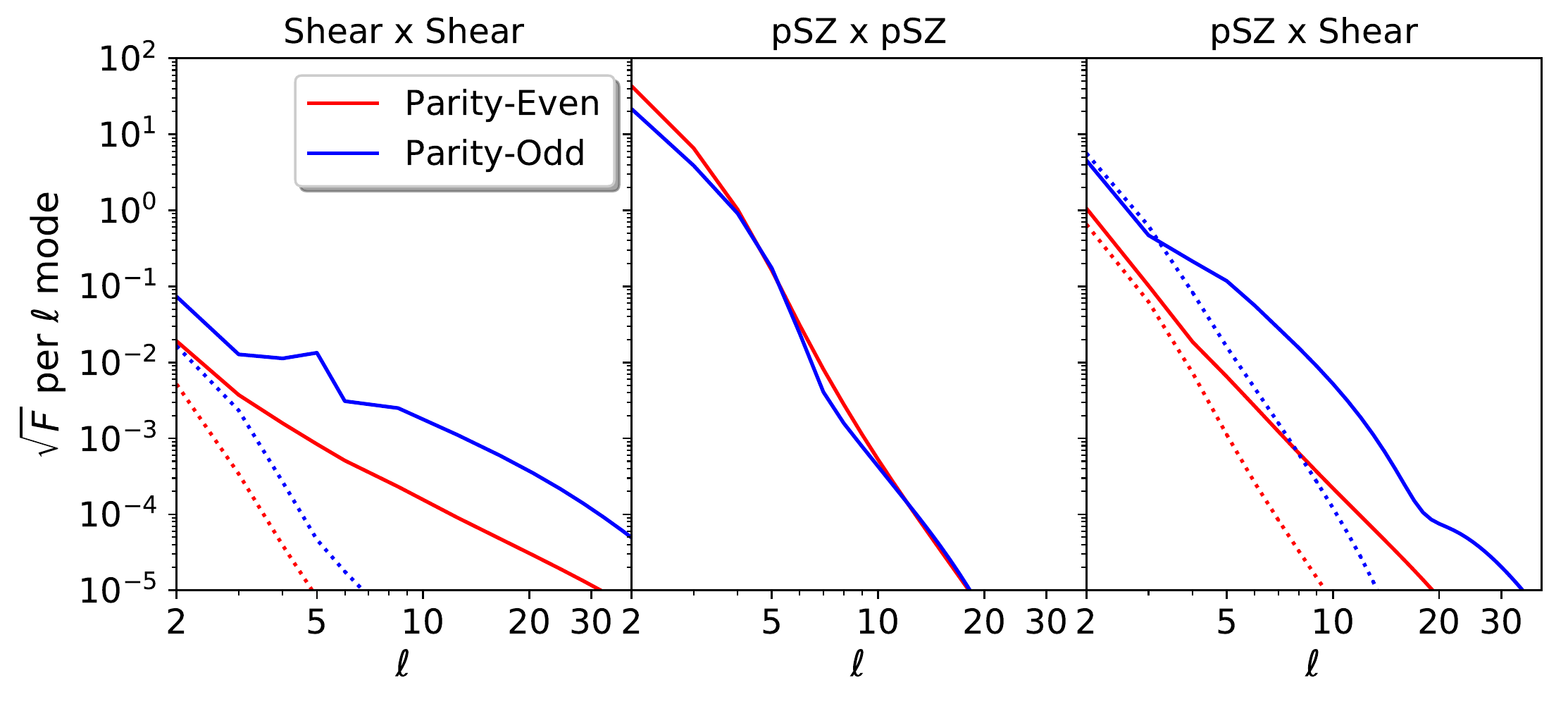}
    \caption{Contributions to the tensor Fisher matrix from shear and pSZ auto- and cross-spectra, assuming CMB-S4/VRO noise parameters. We show results for the parity-even and parity-odd components (constraining $r$ and $r\Delta_h$ respectively) in red and blue, plotting the contribution to $\sqrt{F}=\sigma^{-1}$ from each set of $\ell$ modes. The dashed curves give the results with only lensing shear-modes; we find that lensing dominates the shear constraints, but intrinsic alignments are also important for the cross-spectra with pSZ. $1\sigma$ detection limits are given in Tab.\,\ref{tab: tensor-pSZ}; briefly, we find that the cross-spectrum has some (albeit weak) sensitivity to tensors, particularly parity-odd tensors, and, suffers less from from systematic effects than auto-spectra.}
    \label{fig: SN-tensor}
\end{figure}

The forecasted constraints on $r$ are shown in Tab.\,\ref{tab: tensor-pSZ} and Fig.\,\ref{fig: SN-tensor}. As found previously \citep[e.g.,][]{2012PhRvD..86h3513S}, shear auto-spectra are not able to place tight constraints on tensors: even with the more optimistic noise profile, we find $\sigma(r) = 20$, several orders of magnitude weaker than the current constraints from BICEP \citep{BICEPKeck:2022mhb}. This is partly caused by the VRO galaxy sample, whose source density peaks at $z\sim 0.3$ (albeit with a broad tail); a sample extending to higher $z$ would allow for considerably more stringent limits. 

For the pSZ-auto spectra, we find much tighter constraints, exceeding the current BICEP limits. Whilst these are unlikely to be competitive in the near future, given the rapid advance in CMB detector technology \citep[e.g.,][]{Deutsch:2018umo}, they are nevertheless interesting, since the signal arises from a small scale (doubly-squeezed) $\av{\delta_g^2(Q\pm iU)^2}$ trispectrum rather than the usual $\av{BB}$ signal, and is less subject to lensing and atmospheric effects. For the cross-spectra, we find weaker constraints, with a similar $\ell$-dependence to the auto-spectra. For CMB-S4/VRO, we forecast a $1\sigma$ constraint of $\sigma(r) = 0.9$ (in accordance with \citep{Deutsch:2018umo}), which increases only to $0.2$ with the more optimistic noise profiles of CMB-HD/VRO. This is unlikely to be of use in the near future. Furthermore, the constraint scales with one-power of the CMB noise amplitude, $\Delta_P$ (since the cross-spectrum involves only one polarization field) and thus improves slower than the auto-spectra when the noise is reduced (noting that the galaxy noise primarily arises from cosmic variance at low redshift, though, as before the situation is better at high redshift). Finally, we note that, unlike for shear, the pSZ-shear correlation arises primarily due to lensing effects, rather than intrinsic alignments, and is insensitive to the redshift binning, with $\lesssim10\%$ change to $\sigma(r)$ if the number of bins is increased to 30.

\begin{table}[]
    \renewcommand{\arraystretch}{1.3}
    \caption{$1\sigma$ errorbar on the amplitude of parity-even (left) and parity-odd (right) tensors, $r$ and $r\Delta_h$, from the pSZ and shear data-sets, computed via the Fisher matrix formalism of \eqref{eq: fisher}. Results are shown for two noise parameters, as in Tab.\,\ref{tab: scalar-pSZ}. We consider constraints from both lensing and intrinsic alignment (IA) contributions to the shear signal, which we note give significant cancellation. Odd-parity constraints involving shear are significantly tighter than those for even-parity spectra since the $EB$ spectra are not cosmic-variance limited in linear theory. As in previous work, the constraining power of shear auto-spectra is very weak, but the cross-spectra offer a potential avenue for detecting tensors, albeit in the distant future.}\label{tab: tensor-pSZ}
    \centering
    \subfloat[Even Parity]{
    \begin{tabular}{l|cc|cc|c}
    \hline\hline
    $\sigma(r)$ & $\Delta_P$ & $\bar{n}$ & Lensing & \quad IA \quad & Total\\\hline
    Shear $\times$ Shear & 1 & 40 & 190 & 114 & 51\\
    pSZ $\times$ pSZ & 1 & 40 & $-$ & $-$ & 0.023\\
    pSZ $\times$ Shear & 1 & 40 & 1.5 & 2.0 & 0.94\\\hline
    Shear $\times$ Shear & 0.5 & 100 & 66 & 39 & 17\\
    pSZ $\times$ pSZ & 0.5 & 100 & $-$ & $-$ & 0.0050\\
    pSZ $\times$ Shear & 0.5 & 100 & 0.42 & 0.54 & 0.26\\\hline\hline
    \end{tabular}
    }
    \hskip 16pt
    \subfloat[Odd Parity]{
    \begin{tabular}{l|cc|cc|c}
    \hline\hline
    $\sigma(r\Delta_h)$ & $\Delta_P$ & $\bar{n}$ & Lensing & \quad IA \quad & Total\\\hline
    Shear $\times$ Shear & 1 & 40 & 60 & 21 & 13\\
    pSZ $\times$ pSZ & 1 & 40 & $-$ & $-$ & 0.046\\
    pSZ $\times$ Shear & 1 & 40 & 0.18 & 0.46 & 0.22\\\hline
    Shear $\times$ Shear & 0.5 & 100 & 22 & 7.5 & 4.7\\
    pSZ $\times$ pSZ & 0.5 & 100 & $-$ & $-$ & 0.012\\
    pSZ $\times$ Shear & 0.5 & 100 & 0.050 & 0.15 & 0.055\\\hline\hline
    \end{tabular}
    }
\end{table}

\subsection{Detectability of Parity-Odd Tensors}
Finally, we consider how one may measure the odd-part of the tensor spectrum using the pSZ and shear. In this case, we require the following spectra:
\beq
    C_\ell^{q^Eq^B}\,,\,C_\ell^{\gamma^E\gamma^B}\,,\,C_\ell^{q^E\gamma^B}\,,\,C_\ell^{q^B\gamma^E},
\eeq
each of which is proportional to the parity-odd amplitude $r\Delta_h$. Interestingly, none of the spectra involving $\gamma^B$ suffer from cosmic variance limitations at leading order, since $\av{\gamma^B\gamma^B}$ contains only noise. As such, we may expect the constraints on parity-odd components to be tighter than those on their even brethren.

Fisher forecasts for $r\Delta_h$ are given in the right panel of Tab.\,\ref{tab: tensor-pSZ} and Fig.\,\ref{fig: SN-tensor}. As foretold, the shear-shear and pSZ-shear constraints on odd-parity tensors are significantly (around an order of magnitude) tighter than for even-parity tensors, but similar for the pSZ auto-spectra, since the relevant estimator involves both CMB $E$- and $B$-modes. That said, our conclusions are similar to before: the shear-auto spectra gives weak constraints, with $\sigma(r\Delta_h) = 13$ for CMB-S4/VRO, whilst the pSZ auto-spectra are somewhat tighter ($0.05$ for CMB-S4/VRO noise), though unlikely to be competitive in the near future. For the cross-spectra, we forecast $\sigma(r\Delta_h) = 0.2$ for CMB-S4/VRO, or $0.06$ for CMB-HD/VRO. Whilst this is still weak, it may be interesting from the point of view of systematics, since the associated CMB primary measurements can often be marred by foregrounds. 

\section{Novel cosmological tests}\label{sec:cosmotests}
In the above, we have highlighted the unique properties of the pSZ-shear cross-correlation. In particular, at low-redshifts the cross-spectrum isolates the ISW component of pSZ, whilst at high-redshifts the SW component is picked out (as in Fig.~\ref{fig: cartoon}). Here, we explore the implications of these properties in the context of a toy-model in order to develop some intuition for their utility. 

Consider a primordial potential perturbation consisting of a single mode $\mathbf{k} = k_0 \hz$ with amplitude $A$. Ignoring radial binning, the multipole moments of an observable $X\in \{\delta_g, \gamma^E, q^E \}$ at fixed comoving distance $\chi$ are given by
\beq
a_{\ell m}^X (\chi) = \sqrt{\frac{2\ell + 1}{4 \pi}} A \Delta_{\ell}^X (k_0, \chi)\delta^{\rm K}_{m0}
\eeq
where $\Delta_{\ell}^X (k_0, \chi)$ are the transfer functions defined in Appedix~\ref{appen: transfer}. In the limit of noiseless measurements of each spectra, the ratio of multipole moments becomes a ratio of transfer functions:
\beq
\frac{a_{\ell 0}^X (\chi)}{a_{\ell 0}^Y (\chi)} = \frac{ \Delta_{\ell}^X (k_0, \chi)}{\Delta_{\ell}^Y (k_0, \chi)}
\eeq

\paragraph{SW Contributions}
At high redshift, where ISW can be neglected, the ratio of the galaxy density to the remote quadrupole signal is
\beq
\frac{a_{\ell 0}^{\delta_g} (\chi)}{a_{\ell 0}^{qE} (\chi)} \propto \frac{(k_0 \chi)^2}{j_2(k_0 [\chi_{\rm dec}-\chi])} \times \frac{D_{\Psi}(\eta_0 - \chi)}{\left(2D_\Psi(\eta_{\rm dec})-\frac{3}{2}\right)}
\eeq
This is the product of a geometrical factor (which in the limit $k_0 \ll 1$ reduces to $\chi^2 / ( \chi_{\rm dec}-\chi)^2$) and a ratio of the potential growth functions at very different times, as previously noted. The ratio of the E-mode shear to the remote quadrupole takes a similar form:
\beq
\frac{a_{\ell 0}^{\gamma^E, \rm IA} (\chi)}{a_{\ell 0}^{qE} (\chi)} \propto \frac{(k_0 H_0^{-1})^2}{j_2(k_0 [\chi_{\rm dec}-\chi])} \times \frac{D_{\Psi}(\eta_0 - \chi)}{\left(2D_\Psi(\eta_{\rm dec})-\frac{3}{2}\right)}
\eeq
for intrinsic alignment, and
\beq
\frac{a_{\ell 0}^{\gamma^E, \rm lens} (\chi)}{a_{\ell 0}^{qE} (\chi)} \propto \int_0^\infty \frac{d\chi'}{\chi'} q_a(\chi') \frac{j_\ell(k\chi')}{j_\ell(k\chi)} \frac{(k\chi)^2}{j_2(k[\chi_{\rm dec}-\chi])} \times \frac{D_\Psi(\eta_0-\chi')}{\left(2D_\Psi(\eta_{\rm dec})-\frac{3}{2}\right)}
\eeq
for the lensing contribution. Both these contributions are also a geometrical factor (which is different to that appearing in the galaxy density expression) times a ratio of potential growth functions, with the lensing contributions weighted by comoving distance. Within this toy model, one could in principle measure the ratios above without cosmic variance, mapping out the geometry of the light cone and the potential growth function with arbitrary precision.  

\paragraph{ISW Contributions}
At low redshifts the SW and Doppler terms can be neglected, and the ratio of galaxy density and the remote quadrupole multipoles becomes:
\beq
\frac{a_{\ell 0}^{\delta_g} (\chi)}{a_{\ell 0}^{qE} (\chi)} \propto \frac{(k_0  \chi)^2 D_{\Psi}(\eta_0 - \chi)}{ \int_{\chi}^{\chi_{\rm dec}} d\bar\chi\,j_2(k_0 [\bar\chi-\chi])\partial_\eta D_\Psi(\eta_0 - \bar\chi)} \sim \frac{(k_0  \chi)^2}{j_2(k_0 [\chi_{\rm dec}-\chi])} \times \frac{D_{\Psi}(\eta_0 - \chi)}{D_{\Psi}(\eta_0 - \chi) - D_{\Psi}(\chi_{\rm dec})}
\eeq
which is a geometrical factor multiplying the fractional change in the potential growth function. The analogous ratio for shear and the remote quadrupole is
\beq
 \frac{a_{\ell 0}^{\gamma^E, \rm IA}(\chi)}{a_{\ell 0}^{qE} (\chi)}\propto \frac{( k_0 H_0^{-1} )^2  D_{\Psi}(\eta_0 - \chi)}{ \int_{\chi}^{\chi_{\rm dec}} d\bar\chi\,j_2(k_0 [\bar\chi-\chi])\partial_\eta D_\Psi(\eta_0 - \bar\chi)} \sim \frac{( k_0 H_0^{-1} )^2}{j_2(k_0 [\chi_{\rm dec}-\chi])} \times \frac{D_{\Psi}(\eta_0 - \chi)}{D_{\Psi}(\eta_0 - \chi) - D_{\Psi}(\chi_{\rm dec})}
\eeq
for intrinsic alignment and
\beq
\frac{a_{\ell 0}^{\gamma^E, 
\rm lens} (\chi)}{a_{\ell 0}^{qE} (\chi)} \propto \int_0^\infty \frac{d\chi'}{\chi'} q_a(\chi') \frac{j_\ell(k\chi')}{j_\ell(k\chi)} \frac{(k_0  \chi')^2 D_{\Psi}(\eta_0 - \chi')}{ \int_{\chi}^{\chi_{\rm dec}} d\bar\chi\,j_2(k_0 [\bar\chi-\chi])\partial_\eta D_\Psi(\eta_0 - \bar\chi)}
\eeq
for lensing. Each of these ratios is dependent on the change in the potential growth function, which is sensitive to the properties of dark energy. Within the context of this toy model, it is therefore possible to put strong constraints on the properties of dark energy, which, due to the off-lightcone properties, are free from cosmic variance.

Extending beyond the toy model described above, information about the geometry of the light cone and the potential growth functions has a more complex encoding in the observables. Nevertheless, the above model illustrates the types of novel cosmological tests that may eventually be possible by combining pSZ and shear or density measurements. 

\section{Discussion}\label{sec: discussion}

This work has considered a novel probe of cosmic history: the correlation of the polarized SZ effect with galaxy shear. Unlike most observables, this is not restricted to the lightcone, and can capture interesting physics in both the scalar and tensor sectors, particularly with regards to the ISW effect and parity-odd gravitational waves. Despite significant theoretical and phenomenological appeal, this cross-correlation appears highly challenging to detect. With the forthcoming generation of surveys, a tenuous detection of the scalar signal may just be within reach, but it is unlikely that the signal can be fully exploited in either this decade or the next. That said, the effect's detectability is limited predominantly by CMB detector noise and the availability of high-redshift galaxies, both of which are likely to improve in the future (for example with MegaMapper \citep{Schlegel:2019eqc}, though the high-redshift tail is limited by reionization). We note an important caveat: this work has only considered linear contributions to the pSZ and shear statistics. In the non-linear Universe, higher-order scalar corrections can lead to non-negligible contributions to both $E$- and $B$-mode observables, which may give a fundamental limitation to how well the various signals can be detected. Furthermore, we have ignored the notorious `optical depth degeneracy', relating to the poorly understood connection between the electron and matter distributions \citep[e.g.,][]{Smith:2018bpn}. Whilst this is an important multiplicative uncertainty, the detections considered herein are sufficiently futuristic that one may cautiously hope such problems to be solved by the relevant time, for example using kSZ measurements or cross-correlations with fast radio bursts \citep{Madhavacheril:2019buy}.

Putting the above issues aside, we close by considering the implications of a detection of the pSZ-shear cross-correlation, assuming that one can be made. Perhaps the most appealing feature of cross-correlations is that they suffer from significantly fewer systematic effects than auto-spectra. As mentioned above, the pSZ-shear spectra is a $\av{\gamma\delta(Q\pm iU)}$ bispectrum, and involves only one power of the CMB: as such, a number of physical effects, including detector noise and calibration errors, will be averaged out. In particular, galactic foregrounds and weak lensing provide serious barriers to extracting tensor-modes from the primary CMB: the former contributes only if a residual galaxy selection function couples to the galactic microwave emission, whilst the latter generally cancels in the pSZ signal, due to the structure of the relevant kernel \citep{Deutsch:2018umo}. Although the constraining power on tensor modes from pSZ is weak, the availability of such a constraint may be an important cross-check in the event of a future detection of tensors from the primary CMB.

Secondly, we have shown that the pSZ-shear cross-correlation is dominated by the ISW effect (unlike pSZ auto-spectra), particularly at low redshifts. Although a robust detection remains far-off, its measurement \resub{would provide} direct evidence for dark energy, inducing the time-variation of the Bardeen potentials. Usual ISW constraints arising from the CMB (mostly commonly via cross-correlations \citep[e.g.,][]{Boughn:2004zm,Dupe:2010zs,Planck:2013owu}) are fundamentally limited by cosmic variance, thus, the pSZ-based measurements, which can recover three dimensional modes instead of the usual two dimensional ones, may provide a useful avenue for an eventual high-significance measurement of the properties of dark energy. Finally, the measurement of the SW effect in the pSZ-shear cross-spectra (which is most prominent at high redshifts), would give a unique insight into the Universe's (in)homogeneous evolution. For a lens at $\chi$, the pSZ-shear cross-spectrum measures the following combination of growth rates: $D(\chi,\vx)D(\chi_{\rm dec},\vx)$ (ignoring a geometric prefactor), which can be compared to that of lensing-alone: $D^2(\chi,\vx)$, at some position $\vx$. As mentioned above, a futuristic measurement could, in principle, be used to map the local (off-lightcone) values of $D(\chi,\vx)/D(\chi_{\rm dec},\vx)$ without cosmic variance, allowing novel tests of the Universe's isotropy and homogeneity, for example probing whether the Universe evolves differently in high- and low-density regions.

Although difficult in practice, a measurement of the pSZ-shear cross-correlation could probe a range of new physics, and shed light on new and unexplored features of the cosmological model.

\begin{acknowledgments}
%\footnotesize
We thank Jo Dunkley, Vid Irsic, Blake Sherwin, Kendrick Smith and the participants of the Flatiron SZ workshop for insightful discussions relating to pSZ, as well as Stephon Alexander, Morgane K\"onig, and David Spergel for discussions on galaxy shape statistics. \resub{We are also grateful to the anonymous referee for an insightful report. OHEP is a Junior Fellow of the Simons Society of Fellows} and thanks the Perimeter Institute for supporting a visit within which this work was conceived as well as the Simons Foundation and Institute for Advanced Study for support. MCJ is supported by the National Science and Engineering Research Council through a Discovery grant. This research was supported in part by Perimeter Institute for Theoretical Physics. Research at Perimeter Institute is supported by the Government of Canada through the Department of Innovation, Science and Economic Development Canada and by the Province of Ontario through the Ministry of Research, Innovation and Science.

The authors are pleased to acknowledge that the work reported in this paper was substantially performed using the Research Computing resources at Princeton University which is a consortium of groups led by the Princeton Institute for Computational Science and Engineering (PICSciE) and the Office of Information Technology's Research Computing.
\end{acknowledgments}

\appendix

\section{Transfer functions}\label{appen: transfer}

In this appendix, we collect the transfer functions necessary to compute the signal spectra in Eq.~\eqref{eq:scalar_cov} and~\eqref{eq:tensor_cov}. Starting with the galaxy density defined in Eq.~\eqref{eq:binnedgal}, which is sourced only by scalars at linear order, we have:
\beq
    i^{-\ell}\Delta_{\ell,a}^{\delta_g,\rm S}(k) = - \int_0^\infty d\chi\,n_a(\chi) \ \frac{2a(\chi)k^2}{3H_0^2\Omega_m}b_g(\chi)D_{\Psi}(\eta_0 - \chi)j_\ell(k\chi),
\eeq
Since our primary goal is to compute the SZ-lensing cross correlations, we do not pursue a detailed accounting of redshift space distortions, magnification, or relativistic corrections to the observed number counts. A discussion of these effects and their correlations with SZ effects can be found in~\citep{Contreras_2019}.

Moving to shear, the scalar contribution to the $E$-mode defined in Eq.~\eqref{eq:binnedscalarshear} has the transfer function
\beq
    \Delta_{\ell,a}^{\gamma^E,\rm S}(k) &=& i^{\ell}\sqrt{\frac{(\ell+2)!}{(\ell-2)!}}\left[-\int_0^\infty d\chi\,n_a(\chi)\frac{b_S(\chi)}{2\chi^2}D_\Psi(\eta_0-\chi)j_\ell(k\chi)+\int_0^\infty \frac{d\chi'}{\chi'}\,q_a(\chi')D_\Psi(\eta_0-\chi')j_\ell(k\chi')\right],
\eeq
where the lensing efficiency is $q_a(\chi')\equiv \int_{\chi'}^\infty d\chi\,n_a(\chi)$. The tensor contributions to the $E$- and $B$-mode shear defined in Eq.~\eqref{eq:binnedtensorshear} have the transfer functions for $X\in\{E,B\}$:\footnote{This matches \citep{2012PhRvD..86h3527S,2012PhRvD..86h3513S}, albeit with slightly modified conventions described in \S\ref{subsec: tensor-conventions}.}
\beq
     -i^{-\ell}\Delta_{\ell,a}^{\gamma^X,\rm{T}}(k) &=& -\,\frac{1}{8}\left.\mathcal{O}_X\left[\hat Q_{\rm IA,T}(x)\right]\frac{j_\ell(x)}{x^{2}}\right|_{x=0}D_{\rm T}(k,\eta_0)\\\nonumber 
    &&\,-\,\frac{1}{8}\int_0^\infty d\chi\,n_a(\chi)\mathcal{O}_X\left[\hat Q_{\rm IA,T}(x)\right]\frac{j_\ell(x)}{x^{2}}\left(1-b_{\rm T}(\chi)a^{-2}(\chi)(\partial_\eta^2+aH\partial_\eta)\right)D_{\rm T}(k,\eta_0-\chi)\\\nonumber 
    &&\,+\,\frac{1}{4}\int_{0}^{\infty}\frac{d\chi'}{\chi'}\left(m_a(\chi')\mathcal{O}_X\left[\hat Q_{\rm lens,T,1}(x')\right]+\bar{m}_a(\chi')\mathcal{O}_X\left[\hat Q_{\rm lens,T,2}(x')\right]\right)\frac{j_\ell(x')}{x'^2}D_{\rm T}(k,\eta_0-\chi'),
\eeq
where we define
\beq
m_a(\chi') = \int_{\chi'}^\infty d\chi\,n_a(\chi), \qquad \bar m_a(\chi') = \int_{\chi'}^\infty d\chi\,\frac{\chi'}{\chi}n_a(\chi),
\eeq
and we have $\mathcal{O}_E=\mathrm{Re}, \mathcal{O}_B=\mathrm{Im}$, $x\equiv k\chi, x'\equiv k\chi'$ and $D_{\rm T}$ is the tensor growth factor, defined after \eqref{eq: P-h}. This uses the operators
\beq\label{eq: QTensor}
    \hat Q_{\rm lens,T,1}(x) &=& -\frac{x}{2}\left[x(x^2+14)+2(7x^2+20)\partial_x+2x(x^2+25)\partial_x^2+14x^2\partial_x^3+x^3\partial_x^4\right.\\\nonumber
    &&\,\left.-2i\left(4 + x^2 + 6 x\partial_x + x^2\partial_x^2\right)\right]\\\nonumber
    \hat Q_{\rm lens,T,2}(x) &=& \frac{1}{2}\left[24(x^2+1)+x^4+16x(x^2+6)\partial_x+2x^2(x^2+36)\partial_x^2+16x^3\partial_x^3+x^4\partial_x^4\right],\\\nonumber
    \hat{Q}_{\rm IA,T}(x) &=& \left[12-x^2+8x\partial_x + x^2\partial_x^2\right] + 2ix\left[4+x\partial_x\right],
\eeq
which act on the spherical Bessel functions.\footnote{These, respectively, correspond to $2\,\hat Q_{2}^*(x)$, $2\,\hat Q_3^*(x)$ and $\hat Q_{1}^*(x)$ in the notation of \citep{2012PhRvD..86h3513S}.}

For the remote quadrupole field we follow \citep{Deutsch:2017ybc,Deutsch:2018umo}. The $E$-mode remote quadrupole sourced by scalars is given by\footnote{In full, the pSZ signal contains two effects: (a) contributions arising from the remote quadrupole observed at the galaxy location, and (b) contributions sourced by photon distortions between scattering and the observer. The second set are higher-order effects (since they are unobservable unless the first is also present), and will be neglected herein.}
\beq
    \Delta_{\ell, a}^{q^E,\rm S}(k) = -  \int_0^\infty d\chi\,n_a(\chi) \ 5i^\ell \sqrt{\frac{3}{8}}\sqrt{\frac{(\ell+2)!}{(\ell-2)!}}\frac{j_\ell(k\chi)}{(k\chi)^2}\left[\G_{\rm SW}+\G_{\rm ISW}+\G_{\rm Doppler}\right](k,\chi),
\eeq 
where 
\beq
    \G_{\rm SW}(k,\chi) &=& -\left(2D_\Psi(\eta_{\rm dec})-\frac{3}{2}\right)j_2(k[\chi_{\rm dec}-\chi])\\\nonumber
    \G_{\rm ISW}(k,\chi) &=& -2\int_{\chi}^{\chi_{\rm dec}} d\bar\chi\,\partial_\eta D_\Psi(\bar\eta)j_2(k[\bar\chi-\chi])\\\nonumber
    \G_{\rm Doppler}(k,\chi) &=& \frac{1}{5}kD_v(\eta_{\rm dec})\left(3j_3(k[\chi_{\rm dec}-\chi])-2j_1(k[\chi_{\rm dec}-\chi])\right).
\eeq
The pSZ effect is also sourced by gravitational waves, with the same structure as \eqref{eq: remote-q} but encoding the tensorial contributions to $\Theta$, arising from post-recombination effects, as in the usual CMB. Following a lengthy calculation outlined in \citep{Deutsch:2017ybc,Deutsch:2018umo}, we find the following transfer functions
\beq
    i^{-\ell}\Delta_{\ell,a}^{q^E,\rm T}(k) &=& \int_0^\infty d\chi\,n_a(\chi) \ \frac{5\sqrt{6}}{4}\mathrm{Re}\left[\hat Q_{\rm IA,T}(x)\right]\frac{j_\ell(x)}{x^2}\int_\chi^{\chi_{\rm dec}}d\bar\chi\,\partial_\eta D_T(\bar\eta)\frac{j_2(k[\bar\eta-\eta])}{(k[\bar\eta-\eta])^2}\\\nonumber
    i^{-\ell}\Delta_{\ell,a}^{q^B,\rm T}(k) &=& -\int_0^\infty d\chi\,n_a(\chi) \ \frac{5\sqrt{6}}{4}\mathrm{Im}\left[\hat Q_{\rm IA,T}(x)\right]\frac{j_\ell(x)}{x^2}\int_\chi^{\chi_{\rm dec}}d\bar\chi\,\partial_\eta D_T(\bar\eta)\frac{j_2(k[\bar\eta-\eta])}{(k[\bar\eta-\eta])^2},
\eeq
in terms of the $\hat Q_{\rm IA,T}$ operator of \eqref{eq: QTensor}. 

\section{Correlations with the Kinetic Sunyaev-Zel'dovich Effect}\label{appen: ksz}
In the above, we have considered the correlations between galaxy shear and the polarized SZ effect. It is interesting to ask also if one expects correlations with the \textit{kinetic} SZ effect (kSZ), given that this is observed at much higher signal-to-noise. Much as the pSZ probes a remote quadrupole, the kSZ effect probes a remote dipole, given by
\beq
    v_{\rm eff}(\chi\hn) = \frac{3}{4\pi}\int d\hn\,\Theta(\chi\hn,\hn')(\hn'\cdot\hn),
\eeq
where $\Theta$ is the CMB temperature fluctuation observed at the location of a distant galaxy. The dipole can be estimated by combining the observed CMB temperature with a tracer of the electron density, with the schematic form
\beq
    \widehat{v}_{\ell m,a} = \sum_{\ell_1m_1\ell_2m_2}W^T_{\ell m\ell_1m_1\ell_2m_2}a_{\ell_1m_1}^T\Delta\tau_{\ell_2m_2,a}
\eeq
where $W^T$ is some weight matrix, $a_{\ell m}^T$ are the CMB harmonics and $\Delta\tau_a$ is the Thomson cross section in bin $a$. This probes the combination $\delta_gT$, such that its auto power-spectrum is $\av{\delta_g^2T^2}$ and its cross-spectrum with shear is a $\av{\gamma\delta_gT}$ bispectrum.

\subsection{Formalism}
Following \S\ref{sec: signal}, the remote dipole power spectrum is given by (from \citep{Deutsch:2017ybc}, adapting to our conventions)
\beq
    \left.C_\ell^{vv}(\chi,\chi')\right|_{\rm S} = 4\pi\int_0^\infty d\log k\,\Delta_\ell^{v,\rm S}(k,\chi)\Delta_\ell^{v,\rm S*}(k,\chi')\P_\Psi(k) \qquad (\ell\geq 1).
\eeq
Unlike for the remote quadrupole, there are no tensor contributions at leading order, since $v$ is a spin-1 field. This defines the kernels
\beq
    i^{-\ell}\Delta_\ell^{v,\rm S}(k,\chi) = \frac{1}{2\ell+1}\left[\mathcal{K}_{\rm SW}+\mathcal{K}_{\rm ISW}+\mathcal{K}_{\rm Doppler}\right](k,\chi)\left[\ell j_{\ell-1}(k\chi)-(\ell+1)j_{\ell+1}(k\chi)\right],
\eeq
with
\beq
    \mathcal{K}_{\rm SW}(k,\chi) &=& 3\left(2D_\Psi(\eta_{\rm dec})-\frac{3}{2}\right)j_1(k[\chi_{\rm dec}-\chi])\\\nonumber
    \mathcal{K}_{\rm ISW}(k,\chi) &=& 6\int_{\chi}^{\chi_{\rm dec}}d\bar\chi\,\partial_\eta D_\Psi(\bar\eta)j_1(k[\bar\chi-\chi])\\\nonumber
    \mathcal{K}_{\rm Doppler}(k,\chi) &=& kD_v(\eta_{\rm dec})\left(j_0(k[\chi_{\rm dec}-\chi])-2j_2(k[\chi_{\rm dec}-\chi])\right)-kD_v(\eta)
\eeq
These can be integrated in redshift as before. 

Finally, we require the noise profile of the remote dipole:
\beq
    \frac{1}{\left.C_\ell^{vv}(\chi,\chi')\right|_{\rm noise}} = \frac{\delta_{\rm D}(\chi-\chi')}{2\ell+1}\sum_{\ell_1\ell_2}\frac{\Gamma^{\rm kSZ}_{\ell\ell_1\ell_2}(\chi)\Gamma^{\rm kSZ}_{\ell\ell_1\ell_2}(\chi')}{C_{\ell_1}^{TT}C_{\ell_2}^{\delta_g\delta_g}(\chi,\chi')}
\eeq
or, after binning in redshift,
\beq
    \frac{1}{\left.C_{\ell,ab}^{vv}\right|_{\rm noise}} = \frac{\delta_{\rm K}^{ab}}{2\ell+1}\sum_{\ell_1\ell_2}\frac{\Gamma^{\rm kSZ}_{\ell\ell_1\ell_2,a}\Gamma^{\rm kSZ}_{\ell\ell_1\ell_2,b}}{C_{\ell_1}^{TT}C_{\ell_2,ab}^{\delta_g\delta_g}}
\eeq
with the definition
\beq
    \Gamma^{\rm kSZ}_{\ell\ell_1\ell_1}(\chi) = \sqrt{\frac{(2\ell_1+1)(2\ell_2+1)(2\ell+1)}{4\pi}}\tj{\ell_1}{\ell_2}{\ell}{0}{0}{0}C^{\Delta\tau\delta_g}_{\ell_2}(\chi).
\eeq

\subsection{Forecasts}
To estimate the utility of the kSZ-shear cross-correlation we utilize Fisher forecasts, as in \S\ref{sec: detectability}. Here, only shear correlations are of relevance (from the SW, ISW and Doppler effects), and we show the corresponding detection significances in Tab.\,\ref{tab: scalar-kSZ}. The kSZ auto-spectrum can be detected at high signal-to-noise in future surveys (which is of no surprise, given that it has been detected in current surveys), with strong detections of both the kSZ-shear and kSZ-galaxy cross-correlations also expected. In contrast to the pSZ signal, the kSZ correlators are dominated by the Doppler term (arising primarily from the source's peculiar velocity); this arises from physics \textit{on} the lightcone, and thus does not add new modes of interest. In the auto-spectra, there is a significant contribution from the SW effect, however, this is reduced from the cross-spectra, with a lower signal-to-noise found even than for pSZ. We note that these results are sensitive to the redshift-binning: increasing to 30 tomographic bins (without photometric errors) roughly doubles the signal-to-noise of the auto-spectra, and amplifies the kSZ-galaxy cross-correlation to a value more comparable with the auto-spectrum. All in all, we conclude that the kSZ cross-spectra are not of particular use if one is interested in off-lightcone physics. However, the large Doppler term may be of use in other contexts, for example in breaking the optical depth degeneracy via a joint shear and kSZ $3\times 2$-point analysis.

\begin{table}[]
    \renewcommand{\arraystretch}{1.3}
    \caption{As Tab.\,\ref{tab: scalar-pSZ} but for the kSZ signal. We find that future surveys will be able to detect both kSZ auto- and cross-spectra at high significance, but that these are strongly dominated by the Dopper velocity term, with the other (off-lightcone) contributions being suppressed even relative to the pSZ case. Both auto- and cross-spectra are significantly enhanced when the number of tomographic bins is increased.\label{tab: scalar-kSZ}}
    \centering
    \begin{tabular}{l|cc|ccc|c}
    \hline\hline
    & $\Delta_P$ & $\bar{n}$ & SW & ISW & Doppler & Total\\\hline
    kSZ $\times$ kSZ & 1 & 40 & 4.2 & 0.28 & 440 & 440\\
    kSZ $\times$ Shear & 1 & 40 & 0.39 & 0.48 & 77 & 77\\
    kSZ $\times$ Galaxies & 1 & 40 & 0.24 & 0.41 & 38 & 38\\\hline
    kSZ $\times$ kSZ & 0.5 & 100 & 32 & 2.0 & 3200 & 3200 \\
    kSZ $\times$ Shear & 0.5 & 100 & 1.8 & 2.2 & 350 & 350\\
    kSZ $\times$ Galaxies & 0.5 & 100 & 0.78 & 1.20 & 110 & 110\\\hline\hline
    \end{tabular}
\end{table}

\bibliographystyle{JHEP}
\bibliography{bibtex}%

\end{document}